# Stellar Metallicities and Gradients in the Isolated, Quenched Low-mass Galaxy Tucana

Sal Wanying Fu[1], Daniel R. Weisz[1], Else Starkenburg[2], Nicolas Martin[3,4], Francisco J. Mercado[5], Alessandro Savino[1], Michael Boylan-Kolchin[6], Patrick Côté[7], Andrew E. Dolphin[8,9], Nicolas Longeard[10], Mario L. Mateo[11], Jenna Samuel[6], and Nathan R. Sandford[12]

[1] Department of Astronomy, University of California, Berkeley, Berkeley, CA 94720, USA; swfu@berkeley.edu
[2] Kapteyn Astronomical Institute, University of Groningen, Postbus 800, 9700 AV, Groningen, The Netherlands
[3] Université de Strasbourg, Observatoire astronomique de Strasbourg, UMR 7550, F-67000 Strasbourg, France
[4] Max-Planck-Institut für Astronomie, Königstuhl 17, D-69117 Heidelberg, Germany
[5] Department of Physics and Astronomy, Pomona College, Claremont, CA 91711, USA
[6] Department of Astronomy, The University of Texas at Austin, 2515 Speedway, Stop C1400, Austin, TX 78712-1205, USA
[7] National Research Council of Canada, Herzberg Astronomy and Astrophysics Research Centre, Victoria, BC V9E 2E7, Canada
[8] Raytheon, 1151 East Hermans Road, Tucson, AZ 85756, USA
[9] Steward Observatory, University of Arizona, 933 North Cherry Avenue, Tucson, AZ 85721-0065 USA
[10] Laboratoire dastrophysique, École Polytechnique Fédérale de Lausanne (EPFL), Observatoire, 1290 Versoix, Switzerland
[11] Department of Astronomy, University of Michigan, 311 West Hall, 1085 South University Avenue, Ann Arbor, MI 48109, USA
[12] Department of Astronomy and Astrophysics, University of Toronto, 50 St. George Street, Toronto, ON M5S 3H4, Canada

## Abstract

We measure the metallicities of 374 red giant branch (RGB) stars in the isolated, quenched dwarf galaxy Tucana using Hubble Space Telescope narrowband (F395N) calcium H and K imaging. Our sample is a factor of ∼7 larger than what is available from previous studies. Our main findings are as follows. (i) A global metallicity distribution function (MDF) with $\langle[\mathrm{Fe/H}]\rangle = -1.55^{+0.04}_{-0.04}$ and $\sigma_{[\mathrm{Fe/H}]} = 0.54^{+0.03}_{-0.03}$. (ii) A metallicity gradient of $-0.54 \pm 0.07$ dex $R_e^{-1}$ ($-2.1 \pm 0.3$ dex kpc$^{-1}$) over the extent of our imaging (∼2.5 $R_e$), which is steeper than literature measurements. Our finding is consistent with predicted gradients from the publicly available FIRE-2 simulations, in which bursty star formation creates stellar population gradients and dark matter cores. (iii) Tucana's bifurcated RGB has distinct metallicities: a blue RGB with $\langle[\mathrm{Fe/H}]\rangle = -1.78^{+0.06}_{-0.06}$ and $\sigma_{[\mathrm{Fe/H}]} = 0.44^{+0.07}_{-0.06}$ and a red RGB with $\langle[\mathrm{Fe/H}]\rangle = -1.08^{+0.07}_{-0.07}$ and $\sigma_{[\mathrm{Fe/H}]} = 0.42^{+0.06}_{-0.06}$. (iv) At fixed stellar mass, Tucana is more metal-rich than Milky Way satellites by ∼0.4 dex, but its blue RGB is chemically comparable to the satellites. Tucana's MDF appears consistent with star-forming isolated dwarfs, though MDFs of the latter are not as well populated. (v) About 2% of Tucana's stars have [Fe/H] < −3% and 20% have [Fe/H] > −1. We provide a catalog for community spectroscopic follow-up.

## 1. Introduction

Environment plays a pivotal role in regulating dwarf galaxy star formation. This theoretical understanding emerges from the empirical morphology–density relation that has been established over decades' worth of observations (e.g., Dressler 1980; Giovanelli et al. 1986; Binggeli et al. 1990; Mateo 1998; Bouchard et al. 2009; Geha et al. 2012), where star-forming (e.g., gas-rich and/or with H$\alpha$ emission) dwarf galaxies are found in the field far from massive hosts, and quenched (e.g., gas-deficient) dwarf galaxies tend to be satellites.

As part of the same theoretical and empirical landscape, quenched, low-mass field dwarf galaxies, exceptions to the rule, are quite rare. The origins of the few that do exist are not well understood. Hypotheses thus far have ranged from classifying them as backsplash galaxies that interacted with a more massive host in the distant past (e.g., Teyssier et al. 2012; Santos-Santos et al. 2023), to being the product of complicated interactions from a reionizing UV background (Pereira-Wilson et al. 2023), to forming instead from interactions with the cosmic web (Benítez-Llambay et al. 2013), but detailed studies are needed to assess the validity and prevalence of various processes.

Alongside the Cetus dSph, the Tucana dSph is one of two quenched field dwarf galaxies known in the Local Group (LG), with present-day distances of ∼877 and ∼1345 kpc from the Milky Way (MW) and M31, respectively (Lavery & Mighell 1992), and sufficiently nearby for resolved stellar population studies. Its distance from either of the LG spiral galaxies suggests that it may have evolved largely independently of a more massive host and therefore is not subject to processes such as ram pressure stripping and tidal fields. Aside from its rare status as a quenched field dwarf galaxy, Tucana is also peculiar among analog dwarf galaxies, as its color–magnitude diagram (CMD) contains multiple distinct morphological overdensities (e.g., bifurcated red giant branch, RGB, and horizontal branch, HB) that are not as conspicuous in other LG dwarf galaxies with even more extended star formation histories (SFHs; e.g., Monelli et al. 2010a; Savino et al. 2019).

In all of these cases, the stellar metallicity distribution function (MDF) of a galaxy is essential to studies decoding the astrophysics driving its present-day features (e.g., pre-enrichment, gas accretion). Detailed observations have been conducted in more LG dwarf galaxies (e.g., Kirby et al. 2013; Walker et al. 2016; Pace et al. 2020; Tolstoy et al. 2023), but similarly detailed studies have been challenging in distant dSphs such as Tucana due to few stars being sufficiently bright for efficient spectroscopic observations. The three spectroscopic studies of Tucana to date, all made using the Very Large Telescope (VLT)—Fraternali et al. (2009; VLT/FORS2), Gregory et al. (2019; VLT/FLAMES+GIRAFFE), and Taibi et al. (2020, hereafter T20; VLT/FORS2)—have yielded 52 stars with metallicity measurements down to $g \sim 23.5$ mag. Additional stellar metallicity measurements in Tucana are needed to fully sample its MDF (e.g., extent of the tails, skewness) and accurately infer the astrophysical picture during its star-forming epoch. One approach to expand the sample, which we take in this paper, is to measure metallicities of even fainter stars using a well-tested photometric metallicity method.

Photometric metallicity measurements based on observations of the metallicity-sensitive, near-UV Ca H and K (CaHK) lines have been a long-standing technique used in studies of resolved stellar populations (e.g., Strömgren 1966; Beers et al. 1985; Karaali et al. 2005; Ross et al. 2013). This technique has also seen increasing use in recent years to conduct studies of LG dwarf galaxies and the MW (e.g., Starkenburg et al. 2017; Chiti et al. 2020; Han et al. 2020; Longeard et al. 2021; Fu et al. 2023; Martin et al. 2023). Building off this legacy, we have conducted a Hubble Space Telescope (HST) imaging program to measure photometric metallicities in RGB stars of Tucana as faint as F475W ∼ 25 mag in order to measure a well-populated MDF, establish empirical trends (e.g., gradients), and provide insight into its formation pathways.

This paper is organized as follows. In Section 2, we present our observations and our stellar member selection process. In Section 3, we describe our methods for measuring individual metallicities as well as MDF summary statistics. In Section 4, we present our MDF measurement results, as well as spatial metallicity trends and metallicity in relation to stellar populations. In Section 5, we discuss the implications of our results. We conclude in Section 6.

## 2. Observations

### 2.1. Observations and Data Reduction

We obtain new WFC3/UVIS F395N imaging for Tucana as part of HST GO-16226 (PI: Fu), which has targeted both Cetus and Tucana. Imaging for Tucana was taken between 2022 July 27 and 31 over the course of 12 orbits. We integrated the narrowband filter for a total of 32,268 s, with the target depth achieving a signal-to-noise ratio (S/N) of >10 at the HB, F475W ∼ 25 mag. We perform dithers to remove hot pixels and reject cosmic rays, following the dither pattern from HST GO-10505 (PI: Gallart). We also require that the main science field overlap with archival broadband F475W and F814W imaging from HST GO-10505. Our metallicity inference technique is analogous to those used by the Pristine narrowband CaHK MW survey (e.g., Starkenburg et al. 2017; Martin et al. 2023), and we require additional broadband filters to provide temperature information. In the Pristine-like color space constructed using HST filters, (F475W − F814W) versus F395N − F475W − 1.5 ∗ (F475W − F814W), stars of different metallicity cleanly separate (e.g., Fu et al. 2022, 2023) and enable our principal science case. Additionally, we impose orientation requirements so that our parallel Advanced Camera for Surveys WFC exposures, taken in F475W and F814W, overlap with parallel fields targeted in HST GO-10505. The parallel fields were designed to target scientifically valuable areas in Tucana's galactic halo; we defer analysis of data for this ancillary science case to a future publication.

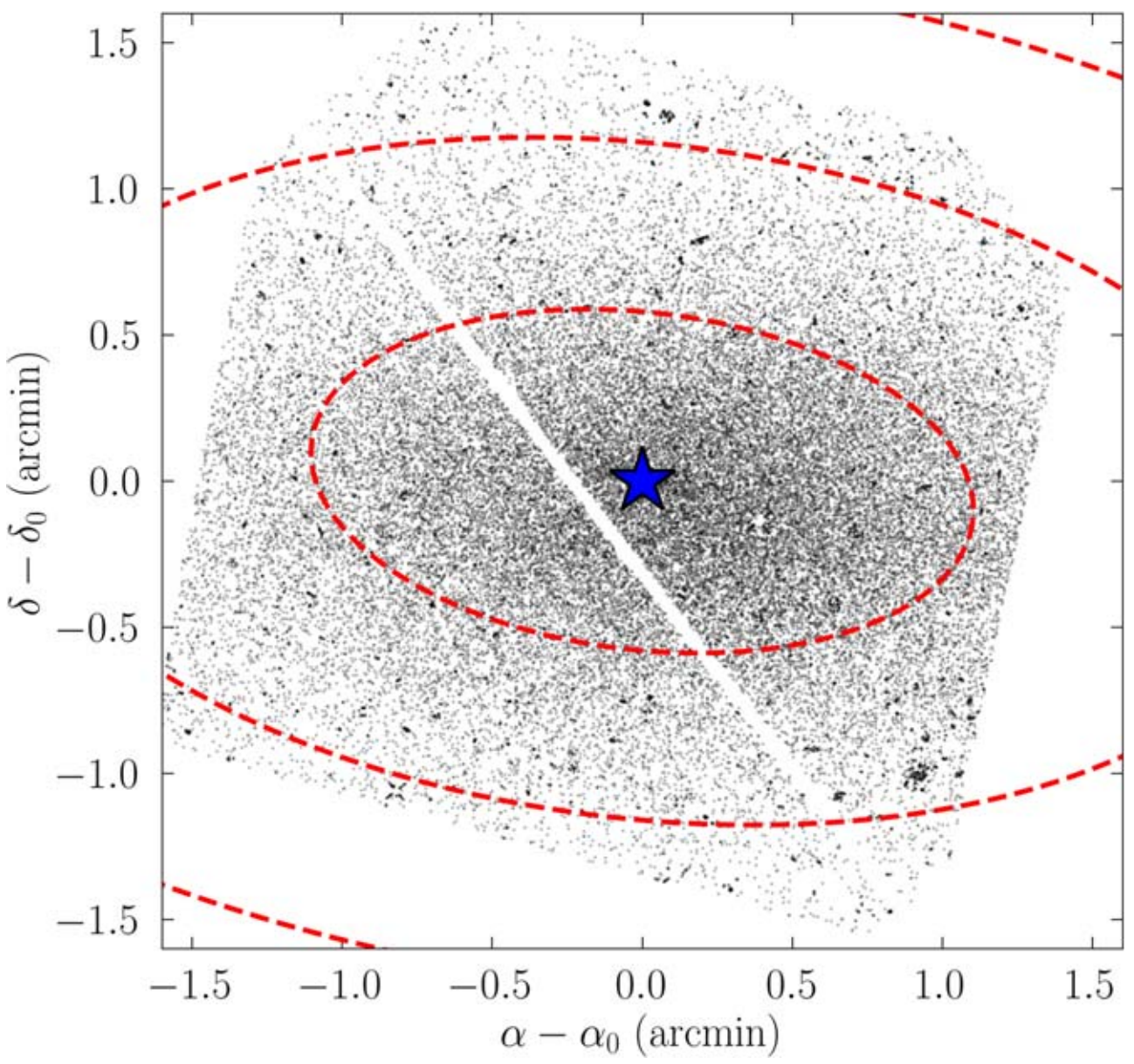

**Figure 1.** FoV of our HST WFC3 F395N imaging compared to the center (blue star) and the 1, 2, and 3 half-light radius contours of Tucana (red ellipses). Small black points are stars that pass our S/N selection criteria. Our data enable stellar metallicity measurements within ∼2.5 $R_e$ of the galaxy, allowing us to measure a global MDF as well as characterize spatial metallicity trends.

We perform point-spread function (PSF) photometry simultaneously on individual F395N, F475W, and F814W `flc` images using DOLPHOT (Dolphin 2000, 2016). We then apply a quality cut on the catalog by requiring S/N > 5, $\text{sharp}^2 < 0.3$, and crowd < 1 in F475W and F814W. We only cull our catalogs using the broadband filters, as they are deeper and therefore more efficient in selecting quality sources.

Figure 1 shows the footprint of our WFC3 imaging compared to the on-sky extent of the galaxy. Small black dots are stars that pass the culling criteria. The blue star shows the center of the galaxy, and the red ellipses trace 1, 2, and 3 $R_e$ of the galaxy (Table 1). Our photometry spans ∼2.5 $R_e$ of the galaxy, allowing us to characterize its spatial metallicity properties in addition to a global MDF (see Section 4.3).

The left panel of Figure 2 shows the resulting broadband CMD for Tucana, constructed from stars that pass the above cuts. Stars with F395N S/N > 3 are color-coded by their respective narrowband S/N. The right panel shows the narrowband CMD of Tucana, overplotted with average narrowband photometric uncertainties as a function of F395N. The high concentration of stars running along the diagonal line faintward of F395N ∼ 25.75 mag are HB stars.

### 2.2. Member Selection

We determine our sample of Tucana member stars by first selecting stars whose colors are consistent with being on the RGB. Most of our stars are too faint to be observed spectroscopically, so we rely primarily on photometric criteria to select the large majority of our sample. In general, we do not expect contamination to have a major impact on our results.

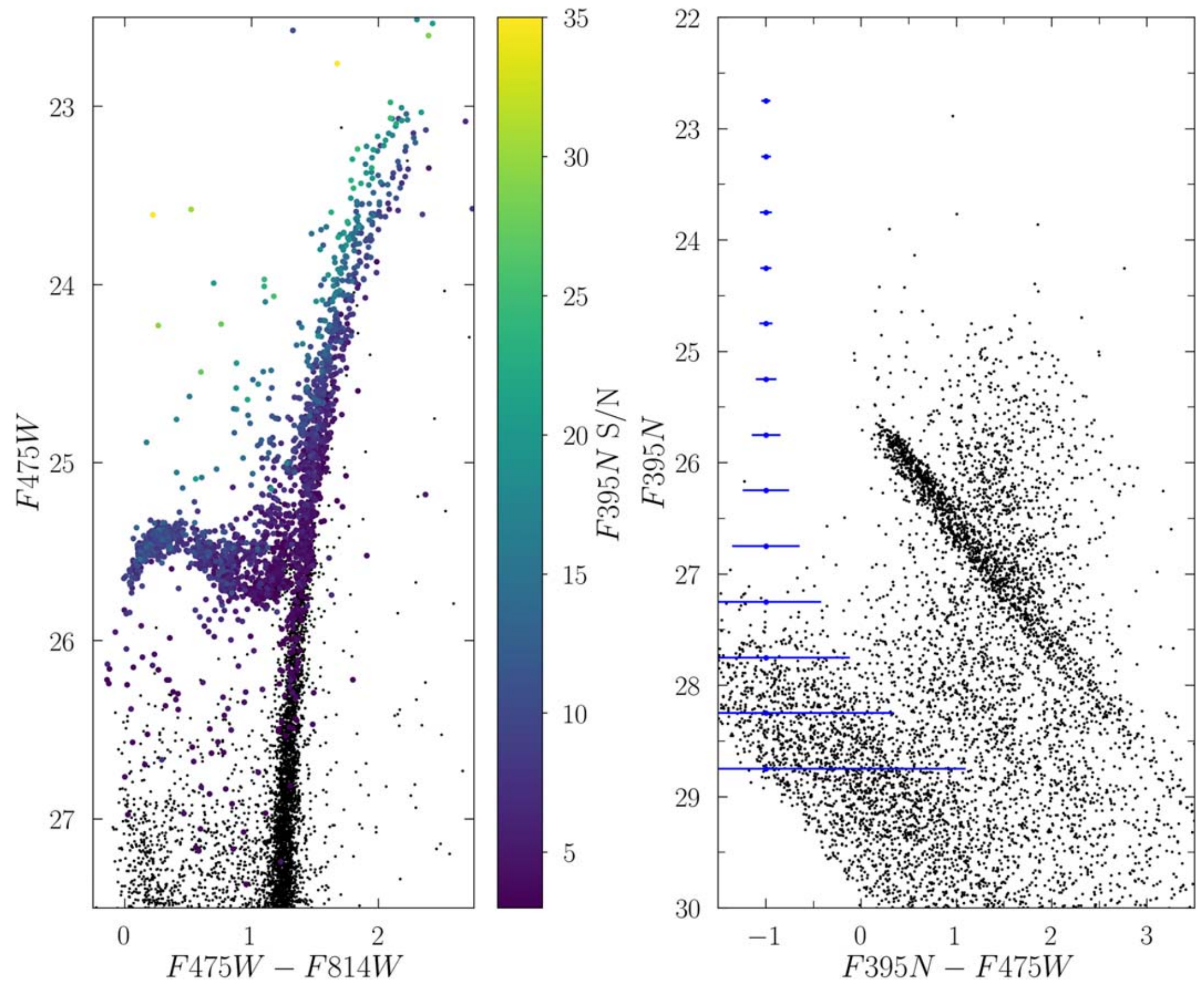

**Figure 2.** Left: the broadband CMD of Tucana. To demonstrate the depth of the narrowband data, we color code stars with F395N S/N > 3 by their narrowband S/N value. Our data reach F395N S/N ∼ 10 at about F475W 24.5 mag, enabling stellar metallicity measurements for stars at least 1 mag fainter than the brightness limit of studies from current ground-based instruments. Right: narrowband F395N CMD of Tucana. The blue error bars show the typical uncertainty as a function of F395N. Uncertainties in F395N − F475W color are driven by uncertainties in F395N. The high concentration of stars running along the diagonal line faintward of F395N ∼ 25.75 mag are HB stars. Stars above the diagonal line are the bright RGB stars that ultimately comprise our analysis sample.

**Table 1**
Tucana Characteristics

| Parameter | Value | Reference |
|---|---|---|
| R.A. (deg) | 340.45667 | Lavery & Mighell (1992) |
| decl. (deg) | −64.41944 | Lavery & Mighell (1992) |
| Ellipticity | 0.48 ± 0.03 | Saviane et al. (1996) |
| P.A. (deg) | 97 ± 2 | Saviane et al. (1996) |
| $R_e$ (arcmin) | 0.8 ± 0.1 | Saviane et al. (1996) |
| Luminosity (Log $L_\odot$) | 5.58 ± 0.01 | Nagarajan et al. (2022) |
| $E(B-V)$ | 0.0268 | Schlafly & Finkbeiner (2011) |
| $(m-M)_0$ | 24.73 ± 0.03 | Nagarajan et al. (2022) |
| $D_\odot$ (kpc) | 886 ± 13 | Nagarajan et al. (2022) |
| $R_e$ in WFC3 FoV | 3.38 | This work |
| F475W exp. time (s) | 34,560 | This work |
| F814W exp. time (s) | 30,976 | This work |
| F395N exp. time (s) | 32,268 | This work |

**Note.** Observational characteristics of Tucana. Information below the horizontal line describes the HST observations used in this work. We use broadband imaging from HST GO-10505 (PI: Gallart). All the HST images used for this work can be found at the following MAST DOI:10.17909/8974-w227.

Due to the small HST WFC3 field of view (FoV), we do not expect enough MW halo stars to fall within it to significantly impact our final measured MDFs (Fu et al. 2023). Additionally, there are no known MW substructures in the vicinity of Tucana found in spectroscopic surveys of the same region of the sky (Fraternali et al. 2009; Gregory et al. 2019; T20) that could introduce contamination at a level above the background expectations from the MW halo.

One selection effect in our data is that at a given magnitude, we will tend to preferentially lose metal-rich (MR) stars in a simple S/N quality cut because more MR stars tend to have lower S/N in F395N as a result of more absorption in the filter. This is similar to the fixed magnitude effect for spectra, e.g., as discussed by Manning & Cole (2017) in the context of the LMC. We also observe this impact in the left panel of Figure 2, where, for example, at F475W ∼ 24, stars on the bluer side of the RGB (which tend to be more metal-poor, MP) have higher S/N than stars on the redder side of the RGB. Thus, for the S/N quality cut, we require that stars in our sample have an S/N greater than 7 in F395N as opposed to the criterion of 10 used in previous CaHK narrowband studies (e.g., Fu et al. 2023). This selection criterion allows us to include most stars brighter than F475W ∼ 24 mag.

For brighter stars in the resulting sample, we then do an additional level of membership vetting by cross-matching our catalog with the data sets analyzed in (T20), which include a reanalysis of Tucana observations from Fraternali et al. (2009) and Gregory et al. (2019). None of the overlapping stars that our catalog has with the spectroscopic studies were ruled out as Tucana radial velocity nonmembers.

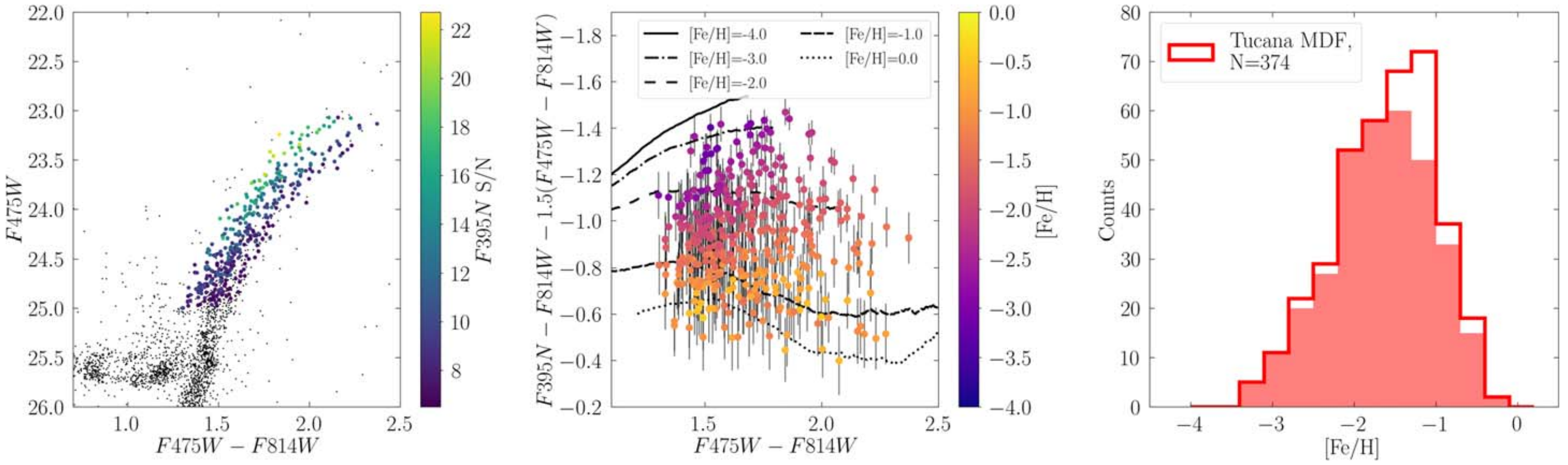

**Figure 3.** Left: selection of Tucana stars along the RGB and their F395N S/N. Center: location of member stars on Pristine-like CaHK space. Overplotted are monometallic CaHK tracks convolved using the AST error profile of a star at F475W = 24.14, the median brightness of stars in our sample. The stars are color-coded by their inferred metallicity, which can also be upper limits on the MP end and lower limits on the MR end. Reasons why the color-coding of the stars does not perfectly follow the trend implied by the monometallic tracks can be attributed to the reporting of measurement limits for ∼50% of stars more MR than [Fe/H] = −1.0 (e.g., Section 4.2), as well as the varied S/N of stars blueward of F475W − F814W = 2.0, which span about 2 mag in brightness. Right: resulting MDF of Tucana. Open histograms represent stars for which we were only able to constrain either an upper limit (on the MP end) or a lower limit (on the MR end). Our data have enabled metallicity measurements for 374 stars down to F475W ∼ 25 mag.

### *2.3. Artificial Star Tests*

We assess the photometric uncertainties for the stars in our sample by running artificial star tests (ASTs). ASTs involve inserting stars of known magnitudes (i.e., in F395N, F475W, and F814W) into the corresponding HST `flc` images and attempting to recover their measurements using the same DOLPHOT PSF-fitting procedure used to obtain the original photometric reduction. By running large numbers of ASTs, we build up the statistics necessary to calculate the photometric uncertainty (scatter between the difference in recovered and input magnitudes) and bias (systematic offsets in the difference in recovered and input magnitudes). As we measure metallicities for individual stars, we do not leverage the completeness information in the ASTs that are used for population-level inferences in, e.g., star formation history (SFH) studies.

Following Fu et al. (2023), we generate ASTs for each potential member star in Tucana that passes our initial color cut. We center the ASTs for each star within 0.2 mag of its F475W magnitude and require that the input ASTs fall within the color space of $0.7 < \mathrm{F475W} - \mathrm{F814W} < 2.5$ and $-2.0 < \mathrm{CaHK} < -0.2$. The 10,000 ASTs we run are distributed to cover all the models in the Pristine-like CaHK color space that we will use to measure individual metallicities.

In Section 3.1, we discuss how we incorporate the results of the ASTs into our individual metallicity inference procedure.

## 3. MDF Measurements

### *3.1. Individual Metallicity Measurements*

Our metallicity measurement method largely follows that from Fu et al. (2023), which measured metallicities in ultra-faint dwarf galaxies (UFDs) using HST CaHK data. For fitting the CaHK data of individual stars in our sample, we adopt a fitting technique that is statistically similar to the long-established CMD SFH modeling approach (Dolphin 2002). Here, we briefly describe this process and provide updates made to the method to better describe the case of a galaxy with an SFH more extended than a UFD.

We first construct the equivalent of a Hess diagram (i.e., a density map) in the Pristine-like color space shown in the center panel of Figure 3 using bins of 0.025 by 0.025 mag. We then convolve our synthetic stellar population models, which have a range of metallicities, by the uncertainty and bias profile as set by our ASTs. Afterward, we infer stellar metallicities by using a Poisson likelihood function to compare the overlap of an individual star's CaHK color–color properties with that of model CaHK tracks of various metallicities that have been corrected for observational effects following the results of the ASTs.

One notable departure from the Fu et al. (2023) method is that here we account for the impact of varying [$\alpha$/Fe] over the range of metallicities considered. In dwarf galaxies of comparable luminosity to Tucana, the expectation is that there will be variations in [$\alpha$/Fe] due to its extended star formation (Monelli et al. 2010a; Savino et al. 2019), specifically, the decline of [$\alpha$/Fe] values at higher metallicities due to the delayed onset of Type Ia supernovae (SNe; e.g., Tinsley 1980; Gilmore & Wyse 1991). We address these issues in our choice of priors for the metallicity inference process.

We begin by using soon-to-be-released v2 models from the MESA Stellar Isochrones and Tracks (MIST) isochrone suite (Choi et al. 2016; Dotter 2016). We choose this set of models over other available models because their grid calculates models for the largest range of metallicities ([Fe/H] = −4.0 to +0.5) and over a range of [$\alpha$/Fe] enhancements ([$\alpha$/Fe] = +0.0 to +0.4). We obtained metallicity grids generated for populations with [$\alpha$/Fe] = +0.0, [$\alpha$/Fe] = +0.2, and [$\alpha$/Fe] = +0.4 to capture the range of $\alpha$ enhancements observed in dwarf galaxies (e.g., Tolstoy et al. 2009). The spacing between the metallicity models is 0.05 dex and ranges from [Fe/H] = −4.0 up to [Fe/H] = +0.0.[13] We then convolve these synthetic models through observational effects, which are (1) dust corrections using the extinction values from Schlafly & Finkbeiner (2011) and filter-specific coefficients from MIST and (2) bias and error effects as characterized by the results of our ASTs described in Section 2.3.

[13] MIST generates models up to [Fe/H] = +0.5, but past [Fe/H] = +0.0, the CaHK models are unable to distinguish between stars of different metallicity. We therefore truncate our grid at +0.0. In practice, we can only obtain lower limit constraints for stars that fall at this boundary.

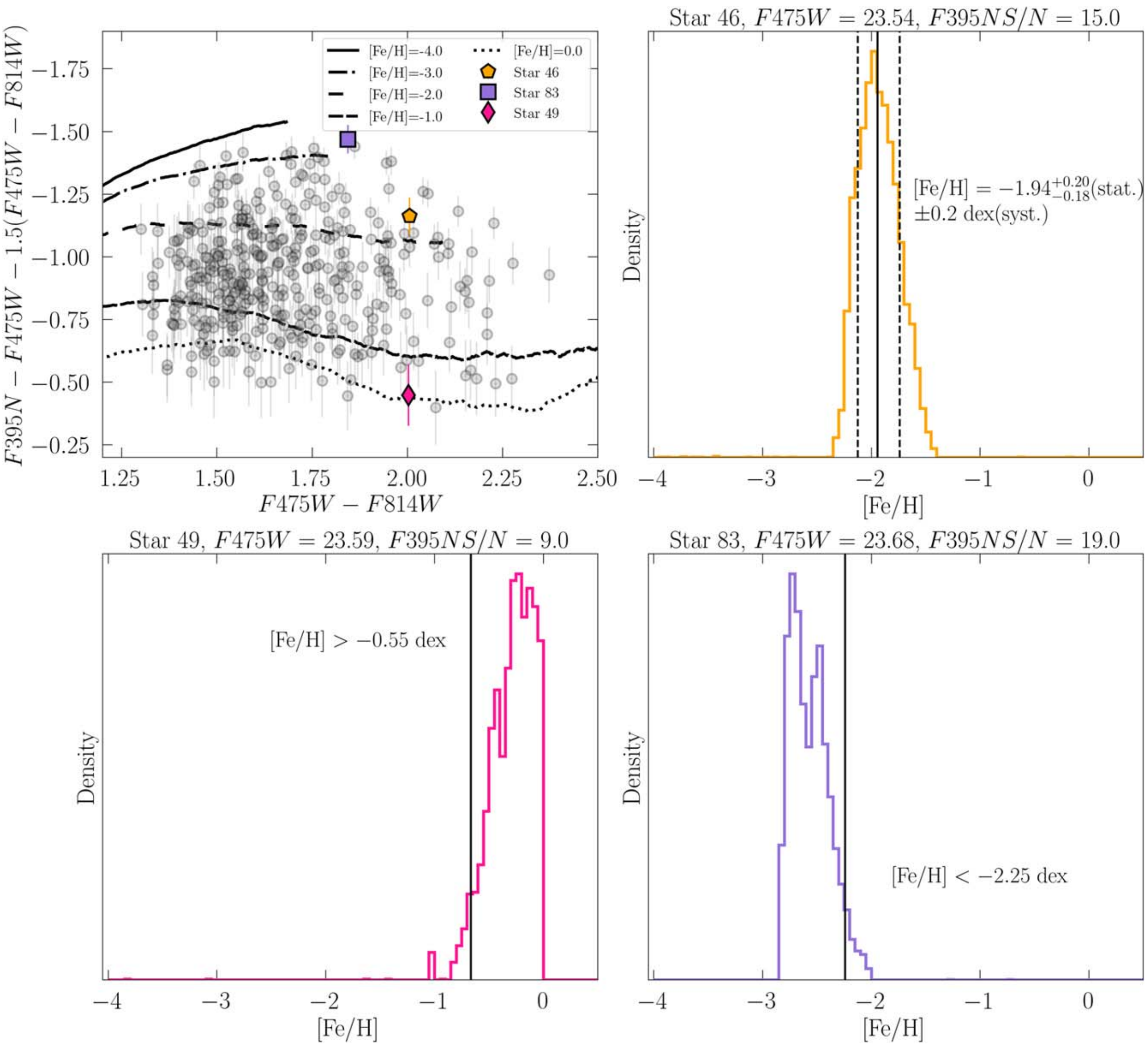

**Figure 4.** Example posterior distributions for examples of stars with a well-constrained measurement (90% of our sample), an upper limit constraint (1% of our sample), and a lower limit constraint (9% of our sample) and their corresponding positions in CaHK color–color space, as well as their corresponding summary statistic. The AST-convolved tracks plotted here are the same as the ones in Figure 3.

To compare the Hess-like diagram of individual stars to our resulting models, we adopt a Poisson likelihood function with the form

$$\log L = \sum_{m_i \neq 0} d \ln m_i - m_i - \ln(d_i!), \quad (1)$$

where $m_i$ are the number of counts in the model bin, and $d_i$ is the data in each bin.

Since there are no constraints on [$\alpha$/Fe] abundances in Tucana from currently available spectroscopy, we adopt priors to account for the impact of varying [$\alpha$/Fe] across metallicity by referring to observed trends for classical dwarf galaxies (e.g., Tolstoy et al. 2009). Our priors are as follows. (1) For metallicity measurements below $[\mathrm{Fe/H}] < -2.0$, we assume $[\alpha/\mathrm{Fe}] = +0.4$. (2) For metallicity measurements above $[\mathrm{Fe/H}] > -1.0$, we assume $[\alpha/\mathrm{Fe}] = +0.0$. (3) For the intermediate range $-2 < [\mathrm{Fe/H}] < -1$, we use the $[\alpha/\mathrm{Fe}] = +0.2$ models to infer metallicities. This choice is to approximate the knee-to-ankle transition observed in [$\alpha$/Fe] versus [Fe/H], although at fixed [Fe/H], there is also a scatter in [$\alpha$/Fe] of ∼0.2 dex. As quantified in Fu et al. (2022), the difference between inferring individual metallicities assuming $[\alpha/\mathrm{Fe}] = +0.4$ versus $[\alpha/\mathrm{Fe}] = +0.0$ is to shift the measurement by about 0.2 dex in the MR direction with lower [$\alpha$/Fe], so the systematic measurement uncertainty we adopt accounts for uncertainty for scatter in [$\alpha$/Fe] at fixed metallicity and uncertainty in the form of the $\alpha$-metallicity relation. In Appendix A, we explore the impact that assuming a different form for the $\alpha$-to-Fe knee pattern has on our resulting MDF.

We then sample the resulting posterior distribution using `emcee` (Foreman-Mackey et al. 2013). We initialize 50 walkers and run the chain for 10,000 steps, with a burn-in time of about 50 steps per star. We monitor convergence using the Gelman–Rubin (GR) statistic (Gelman & Rubin 1992)

Subsequent inferences to calculate MDF summary statistics and metallicity trends (e.g., Section 4.3) assume symmetric,

Gaussian uncertainties on individual star metallicity measurements. In reality, this is not the case because the spacing between monometallic CaHK tracks is uneven and because there are stars for which we can only constrain lower or upper limits (e.g., see Figure 4). We thus follow the methods in Fu et al. (2023) to make the following adjustments to our measurements for population-level inferences.

For stars whose posterior distributions are well constrained enough to measure median and 68% confidence interval uncertainties, we average the uncertainties and add them in quadrature with the 0.2 dex systematic metallicity uncertainty for RGB stars quantified in Fu et al. (2023) to arrive at the final measurement.

For posterior distributions that only constrain either a lower or an upper limit, we adopt measurements that are defined by the median and the 68% confidence interval of the posterior distributions. For stars with upper limits below −3 (i.e., an extremely MP, EMP, star candidate), we add uncertainties in quadrature with a systematic uncertainty of 0.5 dex to account for the reduced sensitivity of CaHK at low metallicities, though these stars are rare in our sample to begin with. For all other cases, we add the uncertainties in quadrature with their corresponding systematic uncertainty, which is already likely to be small in comparison to the posterior-distribution-defined uncertainties.

### 3.2. MDF Summary Statistics

In this section, we describe the procedure for calculating summary statistics of Tucana's MDF for comparison to the literature.

First, we measure the mean and dispersion of the MDF of Tucana by assuming that it is characterized by a single Gaussian distribution. We follow community convention by inferring the MDF using a two-parameter Gaussian likelihood function used by Walker et al. (2006) and subsequent studies of LG dwarf galaxies (e.g., Li et al. 2017; Simon et al. 2020):

$$\log L = -\frac{1}{2}\sum_{i=1}^{N}\log(\sigma^2_{[\mathrm{Fe/H}]} + \sigma^2_{[\mathrm{Fe/H}],i}) - \frac{1}{2}\sum_{i=1}^{N}\frac{([\mathrm{Fe/H}]_i - \langle[\mathrm{Fe/H}]\rangle)^2}{\sigma^2_{[\mathrm{Fe/H}]} + \sigma^2_{[\mathrm{Fe/H}],i}}, \quad (2)$$

where $\langle[\mathrm{Fe/H}]\rangle$ and $\sigma_{[\mathrm{Fe/H}]}$ are the mean metallicity and metallicity dispersion of the galaxy and $[\mathrm{Fe/H}]_i$ and $\sigma_{[\mathrm{Fe/H}],i}$ are the metallicity and metallicity uncertainty for an individual star.

We adopt a uniform prior on the mean and require it to remain within the range set by the most MP and MR stars in Tucana. We also require that the dispersion be greater than or equal to 0. We use `emcee` to sample the posterior distribution, initializing 50 walkers for 10,000 steps. The autocorrelation time is about 50 steps, and we use the corresponding GR statistic to assess convergence.

Next, we compute statistics that quantify higher-order features of the MDF, such as skew and kurtosis, by employing a Monte Carlo method. Similar to the procedure for measuring the mean and dispersion, we assume Gaussian uncertainties on individual metallicity measurements. We use this uncertainty profile to construct 10,000 realizations of the MDF, from which we measure skew and kurtosis. The final measurements we report correspond to the median of the distribution of skew and kurtosis measurements, with lower and upper uncertainties respectively set by the 16th and 84th percentiles.

## 4. Results

In this section, we present the results of our MDF measurements. The center panel of Figure 3 presents the position of our member stars in CaHK space, color-coded by their inferred metallicity. The right panel of the same figure presents the overall MDF. Table 2 presents the MDF summary statistics, along with the sections in which they are discussed. We begin our presentation of the results by discussing individual measurements in Section 4.1 and remark on the overall shape of our MDF in Section 4.2. In Section 4.3, we present our measurement of radial metallicity trends in Tucana, and in Section 4.4, we present the distribution of our metallicity measurements along Tucana's bifurcated RGB. We present our individual measurements as part of Appendix C in Tables 4, 5, and 6, corresponding to the entire sample, the EMP ([Fe/H] $< -3.0$) star candidates, and MR ([Fe/H] $> -1.0$) stars, respectively.

### 4.1. Individual Measurements

The center panel of Figure 3 presents the distribution of stars in our sample in CaHK color space, color-coded by their final inferred metallicity, some of which reflect upper (on the MP end) or lower (on the MR end) limits. Overplotted are a set of representative AST-convolved CaHK monometallic tracks for a star at F475W $\sim$ 24.14, which is the median brightness of stars in our sample. Following the intuition guided by the tracks, the metallicity of stars increases with redder CaHK, but there are some discrepancies in this trend on the MR end of the MDF. This is because stars bluer than F475W − F814W = 2.0 occupy a range of luminosities that is not monotonic in CaHK space. As described in Fu et al. (2023), the bias effects captured by the ASTs are larger for faint stars, and as a result, a faint star would have an inferred metallicity that is more MP than the inferred metallicity of a brighter star occupying a similar position in CaHK color space.

Figure 4 shows examples of individual metallicity posterior distributions to illustrate the nature of our measurements. We have selected three stars that represent the range of posteriors in our fits. As in Fu et al. (2023), most posterior distributions are well within the metallicity limits bounded by our grid and have well-defined peaks. An example is shown in the top right panel of Figure 4. Stars with constrained posterior distributions are often at intermediate metallicities ranging from [Fe/H] = −2.5 to [Fe/H] = −1.0.

Some posterior distributions are truncated at either the MR or the MP end, corresponding to the respective limits of our metallicity grids. We designate measurements as constrained if there is a clear peak in the posterior distribution and if the $1\sigma$ photometric uncertainties of the star fall within the grid. These types of truncated posterior distributions are similar to the ones presented in Figure 5 of Fu et al. (2023), so we do not include them here.

Finally, there are stars on the extremes of the MDF for which we can only constrain either a lower or an upper limit. These stars are at the edge of our grid, and their uncertainties overlap with their respective extreme ends of the grid to permit constraint using the Markov Chain Monte Carlo procedure. We include examples of lower and upper limit posterior

**Table 2**
Tucana MDF Characteristics

| Feature | Parameter | Value | $N$ | Reference in Paper |
|---|---|---|---|---|
| Overall MDF | $\langle$[Fe/H]$\rangle$ | $-1.55^{+0.04}_{-0.04}$ | 374 | Section 4.2 |
| | $\sigma_{\rm [Fe/H]}$ | $0.54^{+0.03}_{-0.03}$ | ⋯ | ⋯ |
| | Skew | $-0.16^{+0.11}_{-0.11}$ | ⋯ | ⋯ |
| | Kurtosis | $-0.02^{+0.28}_{-0.21}$ | ⋯ | ⋯ |
| Gradient within 2.5 $R_e$ | $\nabla_{\rm [Fe/H]}$ (dex $R_e^{-1}$) | $-0.54\pm$ 0.07 | 374 | Section 4.3 |
| | $\nabla_{\rm [Fe/H]}$ (dex arcmin$^{-1}$) | $-0.57\pm$ 0.1 | ⋯ | ⋯ |
| | $\nabla_{\rm [Fe/H]}$ (dex kpc$^{-1}$) | $-2.1\pm$ 0.3 | ⋯ | ⋯ |
| Inner $R_e$ | $\langle$[Fe/H]$\rangle$ | $-1.35^{+0.05}_{-0.05}$ | 203 | Section 4.3 |
| | $\sigma_{\rm [Fe/H]}$ | $0.51^{+0.04}_{-0.04}$ | ⋯ | ⋯ |
| | Skew | $-0.2^{+0.15}_{-0.16}$ | ⋯ | ⋯ |
| | Kurtosis | $-0.04^{+0.44}_{-0.31}$ | ⋯ | ⋯ |
| 1 $R_e < R < 2$ $R_e$ | $\langle$[Fe/H]$\rangle$ | $-1.77^{+0.05}_{-0.05}$ | 160 | Section 4.3 |
| | $\sigma_{\rm [Fe/H]}$ | $0.45^{+0.05}_{-0.05}$ | ⋯ | ⋯ |
| | Skew | $-0.02^{+0.18}_{-0.19}$ | ⋯ | ⋯ |
| | Kurtosis | $0.06^{+0.41}_{-0.30}$ | ⋯ | ⋯ |
| 2 $R_e < R \lesssim 2.5$ $R_e$ | $\langle$[Fe/H]$\rangle$ | $-2.22^{+0.21}_{0.22}$ | 11 | Section 4.3 |
| | $\sigma_{\rm [Fe/H]}$ | $0.54^{+0.27}_{-0.22}$ | ⋯ | ⋯ |
| | Skew | $-0.19^{+0.44}_{-0.44}$ | ⋯ | ⋯ |
| | Kurtosis | $-0.80^{+0.68}_{-0.43}$ | ⋯ | ⋯ |
| Red RGB | $\langle$[Fe/H]$\rangle$ | $-1.08^{+0.07}_{-0.07}$ | 69 | Section 4.4 |
| | $\sigma_{\rm [Fe/H]}$ | $0.42^{+0.06}_{-0.06}$ | ⋯ | ⋯ |
| | Skew | $0.08^{+0.18}_{-0.18}$ | ⋯ | ⋯ |
| | Kurtosis | $-0.70^{+0.31}_{-0.24}$ | ⋯ | ⋯ |
| | $\nabla_{\rm [Fe/H]}$ (dex $R_e^{-1}$) | $-0.42 \pm 0.14$ | ⋯ | ⋯ |
| Blue RGB | $\langle$[Fe/H]$\rangle$ | $-1.78^{+0.06}_{-0.06}$ | 103 | Section 4.4 |
| | $\sigma_{\rm [Fe/H]}$ | $0.44^{+0.07}_{-0.06}$ | ⋯ | ⋯ |
| | Skew | $0.21^{+0.20}_{-0.21}$ | ⋯ | ⋯ |
| | Kurtosis | $0.02^{+0.39}_{-0.31}$ | ⋯ | ⋯ |
| | $\nabla_{\rm [Fe/H]}$ (dex $R_e^{-1}$) | $-0.62 \pm 0.11$ | ⋯ | ⋯ |

**Note.** Summary of metallicity properties of Tucana's MDF from this work. For each property, we include the number of stars used in the calculation and the sections where these results are presented and discussed. The number of stars in the blue and red RGBs do not add up to the total number of stars in our sample because we include an additional magnitude cut in our selection.

distributions in the lower left and lower right panels of Figure 4, respectively.

In general, posterior distributions at the MR end tend to be broader than the posterior distributions for measurements at intermediate metallicities due to (1) the MR stars having a lower S/N on average than the MP stars at a given magnitude and (2) the CaHK tracks having less discriminatory power over different metallicities past [Fe/H] = −0.5.

Next, we compare our measurements to literature metallicities from T20. T20 is the most comprehensive spectroscopic study of Tucana so far. They obtain new observations of Tucana using VLT/FORS2, targeting the near-IR range covering the Ca II triplet lines, and derive metallicities based on the Starkenburg et al. (2010) CaT EW calibration. Alongside their new data, they also analyze previous spectra of Tucana RGB stars from Fraternali et al. (2009; VLT/FORS2) and Gregory et al. (2019; VLT/FLAMES/GIRAFFE). Metallicity measurements were made from the new data set of (T20), referred to in the paper as P91, and from the Fraternali et al. (2009) data set, referred to in the paper as P69. Between the P91 and P69 data sets, we have the largest number of stars in common with the P91 data because its observations are more centrally concentrated on Tucana and have more overlap with our FoV.

We present our comparisons with T20 in Figure 5. The left panel shows where our stars are located on the HST CMD within our WFC3 FoV, as well as where the stars have also been observed with spectroscopy. The center panel of Figure 5 shows one-to-one comparisons between our measurements and those of T20. In total, we have 35 stars in common with this study, with 24 that have metallicity measurements for which we can make direct comparisons. Among the 24 stars, one is a variable star, rendering its CaHK metallicity unreliable,[14] so we exclude it from comparison. In total, we have 23 stars for metallicity comparisons with T20. The measurements show a strong correlation and are in agreement at $\sim 1\sigma$ for constrained measurements (filled circles).

[14] Martin et al. (2023) and references therein discuss how the metallicity of a star can be erroneously inferred if the photometry used to calculate its CaHK color index is taken at different points in its variable phase. As available data are insufficient for characterizing the variability cycle of this star, resolving this issue is beyond the scope of our work.

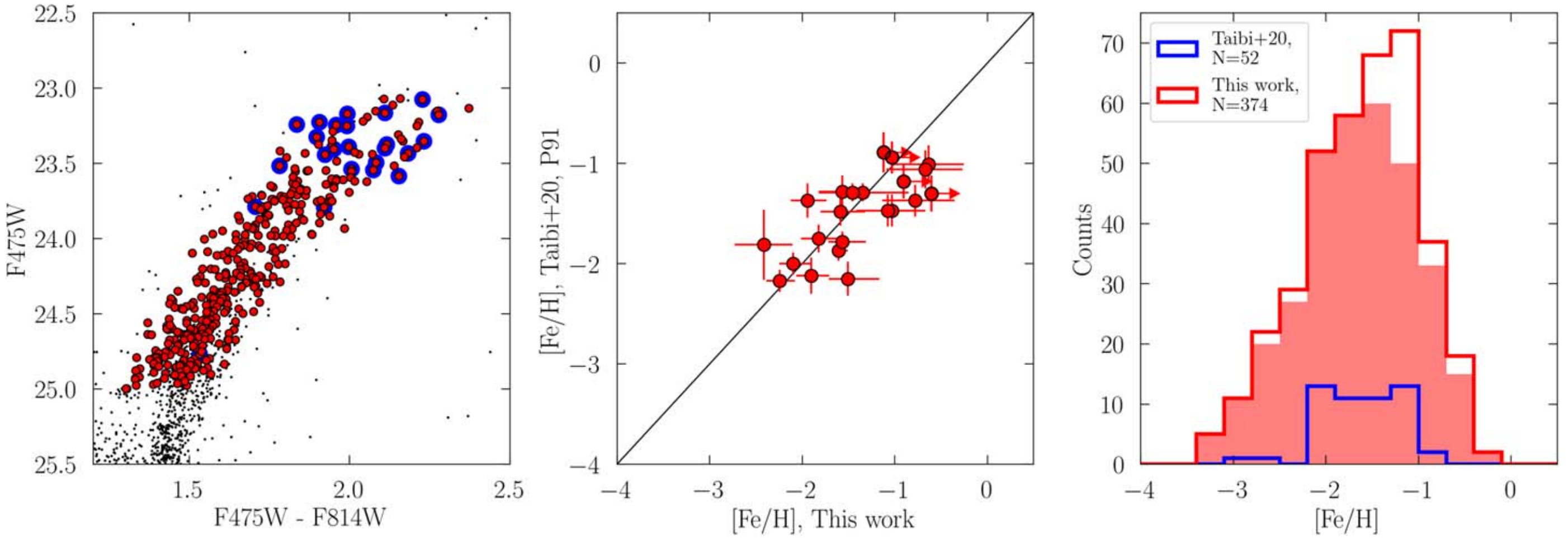

**Figure 5.** Comparing our metallicity measurements to the most comprehensive ground-based CaT EW spectroscopic study of Tucana to date (T20; VLT/FORS2). Left: HST broadband CMD that shows our stars (red) and the stars we have in common with (T20; blue). Center: one-to-one comparison of our metallicity measurements with the measurements that we have in common with (T20). Right: overall MDF comparisons between our metallicity measurements with the MDF from (T20). Our imaging technique permits recovery of metallicities at a similar level of fidelity to state-of-the-art ground-based spectroscopic calibrations, with vastly better sampling along the LF of Tucana to fill out the tails of the MDF.

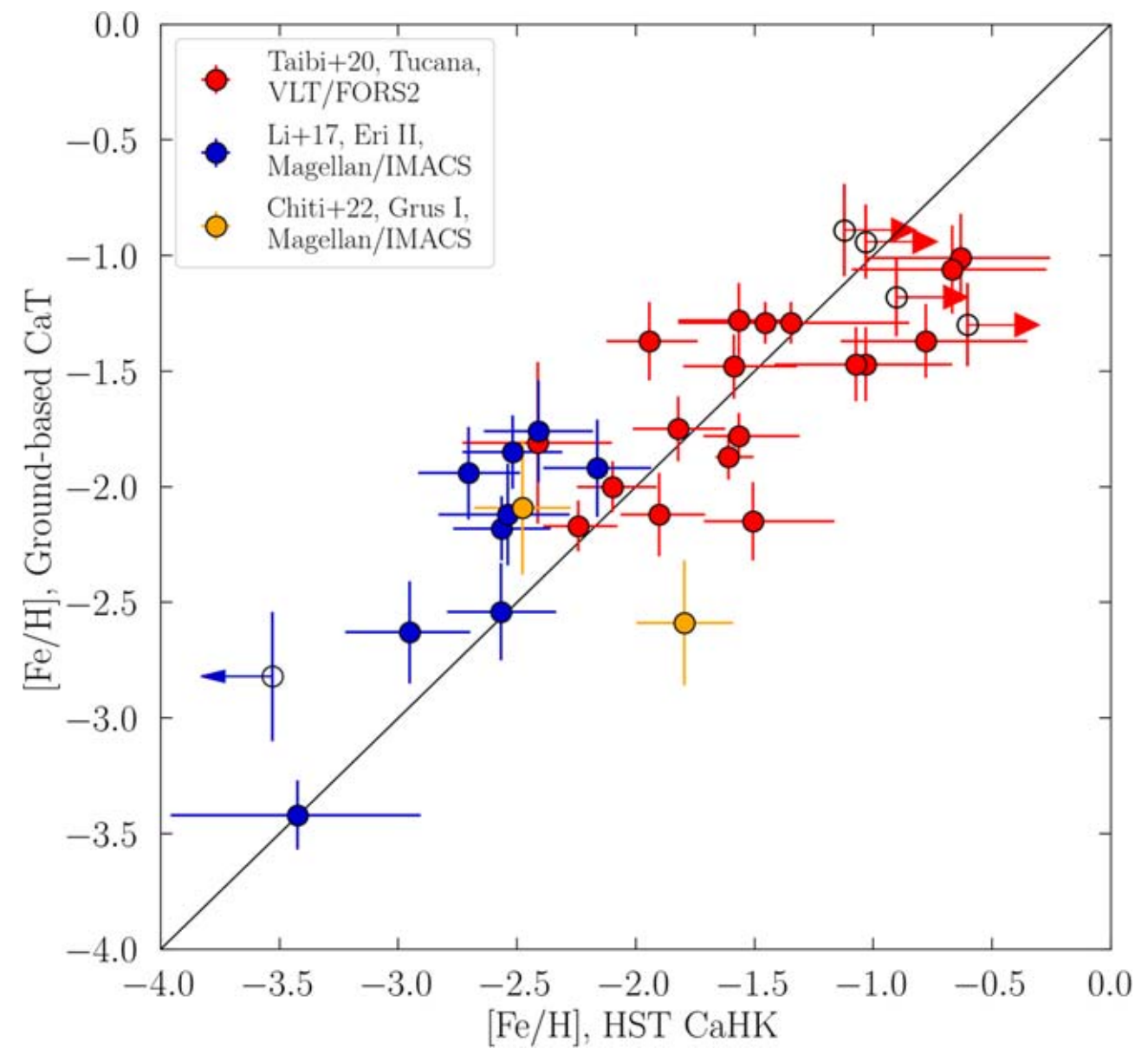

**Figure 6.** Comparing metallicity measurements made using HST CaHK narrowband photometry (UFDs Eri II and Grus I, Fu et al. 2022, 2023; Tucana, this work) against those made using ground-based spectroscopy relying on the CaT EW calibration (Li et al. 2017; T20; Chiti et al. 2022). We include a one-to-one line for visual reference. The measurements show remarkable agreement and demonstrate that CaHK narrowband imaging is a competitive method for measuring stellar metallicities compared with the most commonly used method for LG dwarf galaxies.

To more broadly understand the reliability of CaHK, we now combine all HST CaHK measurements with the literature in Figure 6. We compare photometric stellar metallicity measurements using HST narrowband CaHK imaging (UFDs Eri II and Grus I, Fu et al. 2022, 2023; Tucana, this work) against measurements of the same stars made using the CaT EW calibration from ground-based spectroscopy (e.g., Li et al. 2017; T20; Chiti et al. 2022), with the usual caveats that the CaT methods shown here are heterogeneous (e.g., different calibrations and data quality). Additionally, our CaHK measurement uncertainties reported here also account for systematic uncertainty introduced, e.g., by model and abundance uncertainty (Fu et al. 2023), whereas the measurements reported from CaT EW measurements in these studies include only statistical uncertainties, not ones that may be introduced from different CaT EW calibrations (e.g., Starkenburg et al. 2010; Carrera et al. 2013).

In any case, these measurements show remarkably broad agreement over a large range of metallicities with the long-standing CaT EW method for measuring stellar metallicities in RGB stars of LG dwarf galaxies. Among the constrained measurements, there is no systematic offset, and they are on average within $\sim 1.2\sigma$ agreement; the intrinsic dispersion in this relation is $\sim 0.3\sigma$. A clear takeaway from this comparison is that CaHK imaging provides a robust approach to constructing a large sample of reliable, resolved star metallicities at magnitudes that are often inaccessible to any other current facility. We provide all of our individual measurements in Table 4.

### *4.2. Overall MDF*

We present our global MDF of Tucana in the right panel of Figure 3, constructed from 374 stars. The MDF spans a metallicity range from [Fe/H] = −3.5 to [Fe/H] = −0.5. A total of 60% of the stars are between [Fe/H] = −2.0 and [Fe/H] = −1.0. We identify eight stars with [Fe/H] < −3.0 as EMP star candidates. On the MR end, [Fe/H] > −1.0, we identify 76 stars, with 37 of them being lower limits. The MDF may be slightly skewed negative, as it visually appears to have a slightly longer MP tail, though this may also be in part due to larger uncertainties for MP stars. We present our EMP candidates and MR stars in Tables 5 and 6, respectively.

Table 2 quantifies basic properties of the MDF. For the global MDF, we measure $\langle[\mathrm{Fe/H}]\rangle = -1.55^{+0.04}_{-0.04}$ and $\sigma_{[\mathrm{Fe/H}]} = 0.54^{+0.03}_{-0.03}$. We measure a skew and kurtosis of $-0.16^{+0.11}_{-0.11}$ and $-0.02^{+0.28}_{-0.21}$, respectively. The skew is mildly negative, confirming visual impressions, but is consistent with 0 at $\sim 1.5\sigma$. The kurtosis is consistent with 0, which is also consistent with a clear single peak. Overall, the MDF is well described by a Gaussian distribution.

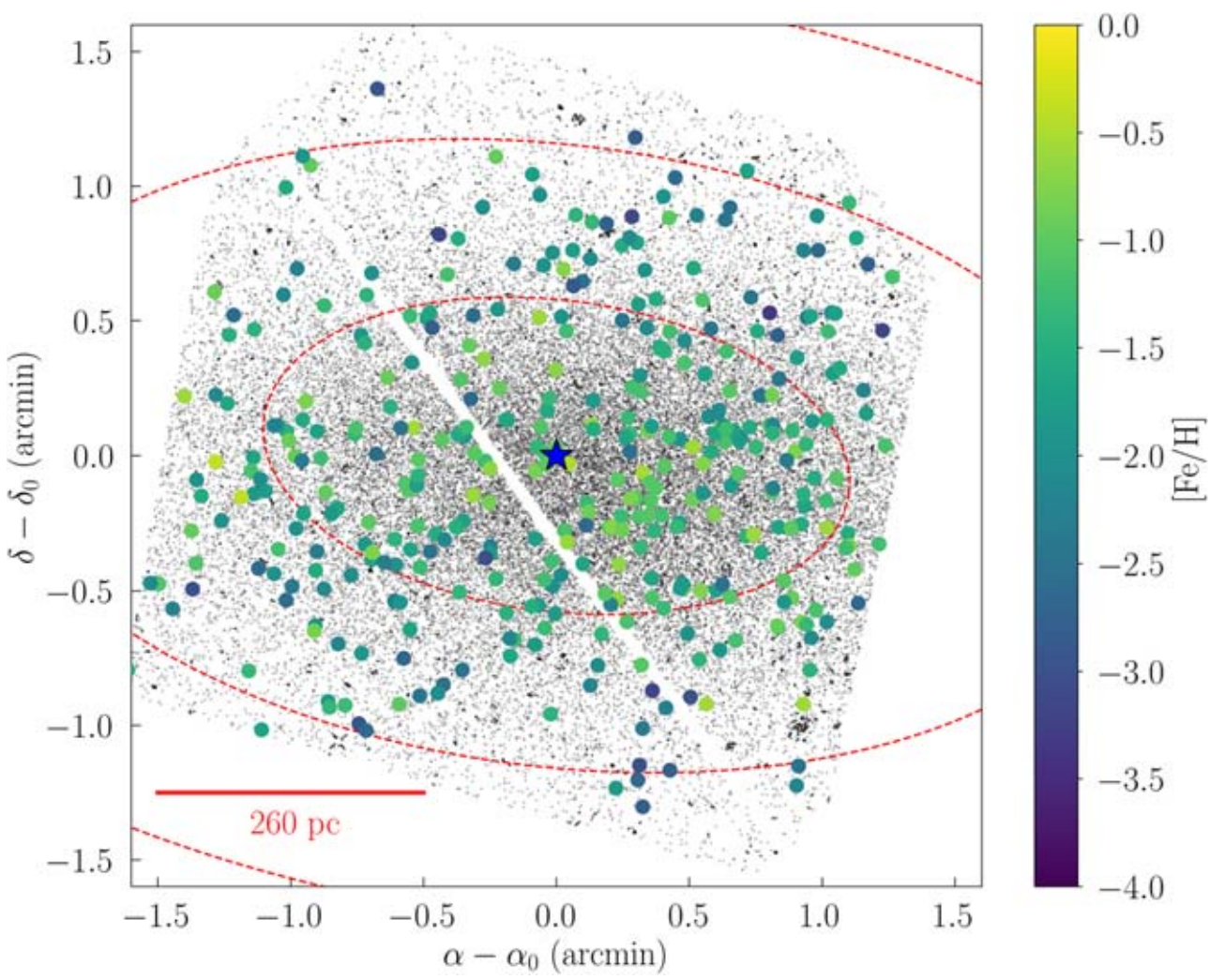

**Figure 7.** The spatial distribution of our CaHK-based stellar metallicities for Tucana. Overplotted are the elliptical half-light radii at 1 $R_e$, 2 $R_e$ and 3 $R_e$. Visually, it is clear that the central region of Tucana has more MR stars and the outer regions have more MP stars.

The blue histogram in the right panel of Figure 5 shows the MDF reported by T20. From 52 RGB stars, T20 measure $\langle[\mathrm{Fe/H}]\rangle = -1.58$ dex and $\sigma_{[\mathrm{Fe/H}]}(\mathrm{intrinsic}) = 0.39$ dex. Using their measurements, we reinfer their MDF properties following the procedure in Section 3.2, obtaining $\langle[\mathrm{Fe/H}]\rangle = -1.64 \pm 0.06$ dex and $\sigma_{[\mathrm{Fe/H}]}$=$0.41 \pm 0.05$ dex. Our mean metallicity is in agreement with the T20 MDF, but we measure a larger metallicity dispersion. On the MP end, 2% of our stars are MP, whereas T20 do not find any stars with $[\mathrm{Fe/H}] < -3.0$. On the MR ($>-1.0$) end, 20% of our stars are MR, while 4% (two stars) from the MDF of T20 are MR. Stars between $[\mathrm{Fe/H}] = -2.0$ and $[\mathrm{Fe/H}] = -1.0$ make up 79% percent of the T20 sample, whereas they are a smaller fraction of our MDF.

We find a larger dispersion than T20 because the CaHK method provides a larger, more complete sampling of the MDF. It allows us to measure metallicities for a nearly complete sample of stars to a faint magnitude limit (F475W $\sim$ 24). No current spectroscopic facility can reach such faint stars, and most spectroscopic observations are challenged by selection effects (e.g., slitmask placement). This MDF comparison illustrates one of the strengths of photometric metallicities.

### 4.3. Radial Trends in Metallicity

The spatial extent probed by our imaging ($\sim$500 pc; $\sim$2.5 $R_e$) and large sample of metallicities allows us to investigate radial trends. Figure 7 shows the spatial distribution of stellar metallicities in Tucana. It is visually apparent that stars in the center of Tucana are higher metallicity than those in the outskirts.

We reinforce this impression by plotting metallicities for each star as a function of radius, along with fits to the data, in Figure 8. We quantify the strength of this gradient first by fitting a line to individual stellar metallicities as a function of elliptical $R_e$ from the center. Following the procedure outlined in Hogg et al. (2010), we adopt their likelihood function, which assumes Gaussian uncertainties on our measurements. Our parameters of interest are the slope $\nabla_{[\mathrm{Fe/H}]}$ and intercept $[\mathrm{Fe/H}]_0$. Additionally, we marginalize over an additional parameter $f$ that quantifies the fractional underestimation of measurement uncertainty. We assume uniform priors for the slope ($-5$ dex arcmin$^{-1}$ $< \nabla_{[\mathrm{Fe/H}]} < 0.5$ dex arcmin$^{-1}$), intercept ($-4$ dex $< [\mathrm{Fe/H}]_0 < 0.0$ dex), and logarithmic fractional uncertainty ($-10.0 < \log f < 1.0$). We sample the resulting posterior distribution using `emcee`. We run 32 walkers for 10,000 steps, with a burn-in time of 50 steps. We monitor convergence using the GR statistic.

Using the entire data set, we measure a metallicity gradient across 2.5 $R_e$ of Tucana of $\nabla_{[\mathrm{Fe/H}]} = -0.54 \pm 0.07$ dex $R_e^{-1}$. For completeness, we also fit a gradient as a function of angular distance from the galaxy, which assumes circular radii from the center. We find a gradient of $\nabla_{[\mathrm{Fe/H}]} = -0.57 \pm 0.1$ dex arcmin$^{-1}$, which translates in physical units to $-2.1 \pm 0.3$ dex kpc$^{-1}$. We report the gradient in units of kiloparsecs to maintain consistency with how gradients in the literature have been reported (Taibi et al. 2022), but because the resulting value is large, its interpretation requires additional orienting remarks. In particular, since the radial extent of Tucana probed by our data is $\sim$500 pc, we caution against interpreting the gradient by extrapolating beyond what is covered by the data (e.g., it may flatten out at larger radii). Within the extent of our radial coverage, the value of the gradient is consistent with the statement that the difference between the center and 2.5 $R_e$ of Tucana is $\sim$1 dex, which we see borne out visually in Figure 8.

In Appendix B.1, we present the correlation plot for the gradient fit using elliptical $R_e$, which we report as the main result for this paper. In Appendix B.2, we present the linear fit and corresponding correlation plots for the inference assuming circular arcminutes, which are comparable to the gradient measured using elliptical $R_e$.

Our large sample size also allows us to measure MDFs in spatial bins across Tucana. We compute the MDFs and associated summary statistics for the MDFs in three spatial bins: $<$1 $R_e$, between 1 and 2 $R_e$, and from $>$2 $R_e$ to the spatial limit of our imaging. We compute mean metallicity, dispersions, skew, and kurtosis for the MDF following the procedure outlined in Section 3.2. The mean metallicity decreases as a function of increasing radius, with $\langle[\mathrm{Fe/H}]\rangle = -1.35^{+0.05}_{-0.05}$, $-1.77^{+0.05}_{-0.05}$, and $-2.22^{+0.21}_{-0.22}$ across the three respective bins. The metallicity dispersions in all three bins have a similar value of $\sim$0.5 dex. We also compute the skew and kurtosis, though both are consistent with zero at $<2\sigma$.

### 4.4. Metallicities along the Split RGB

Distinct stellar populations have long been noted in Tucana as overdensities in various features of its CMD. Harbeck et al. (2001) and Monelli et al. (2010a) note distinct populations in the RGB bump of Tucana, with the latter remarking that similar features are not observed in Cetus, a dSph with similar environmental and SFH properties. Savino et al. (2019) similarly note multiple CMD overdensities in the HB of Tucana, pointing to the presence of distinct populations.

As highlighted in Figure 9, these features manifest among the brightest stars in Tucana as a visual bifurcation in the RGB. We characterize the metallicity properties of the respective bifurcations by selecting RGB stars that are brighter than F475W $\sim$ 24.2 mag. Above this magnitude, the separation is most visually apparent, and we avoid the completeness issues with the preferential exclusion of MR stars toward the fainter end of our sample. We then fit the MDFs of each RGB following the procedure described in Section 3.2. The resulting MDFs with their summary statistics are shown in the right panel of Figure 9. For the blue RGB MDF, we measure $\langle[\mathrm{Fe/H}]\rangle = -1.78^{+0.06}_{-0.06}$ and $\sigma_{[\mathrm{Fe/H}]} = 0.44^{+0.07}_{-0.06}$. For the red RGB MDF, we measure $\langle[\mathrm{Fe/H}]\rangle = -1.08^{+0.07}_{-0.07}$ and $\sigma_{[\mathrm{Fe/H}]} = 0.42^{+0.06}_{-0.06}$. Our measurements unambiguously show that metallicity is at least partially responsible for the RGB split.

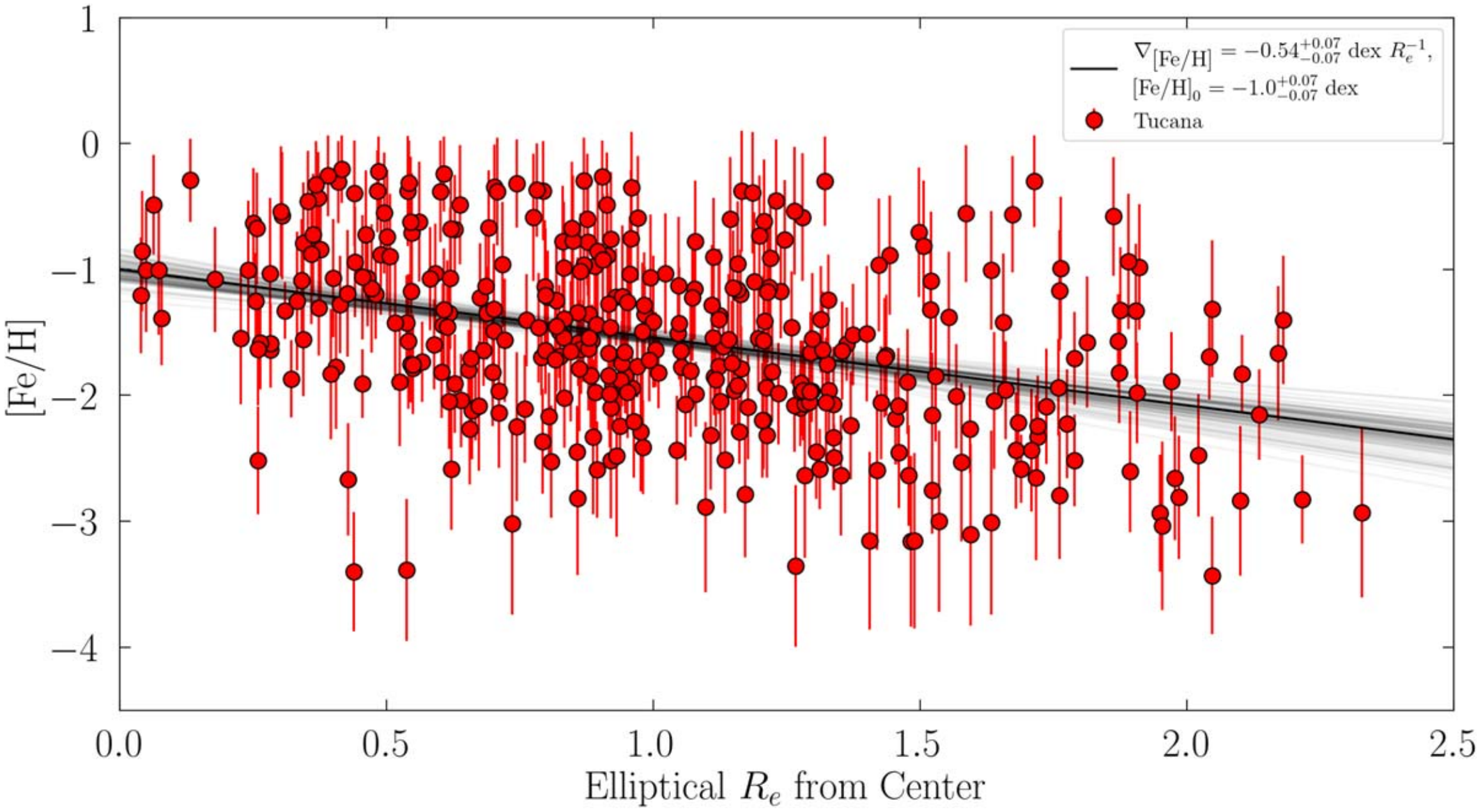

**Figure 8.** Result of fitting a linear model to our Tucana data as a function of elliptical $R_e$ from its center. From our well-sampled MDF, which gives a clear shape to the MP and MR tails, we are able to robustly recover a metallicity gradient.

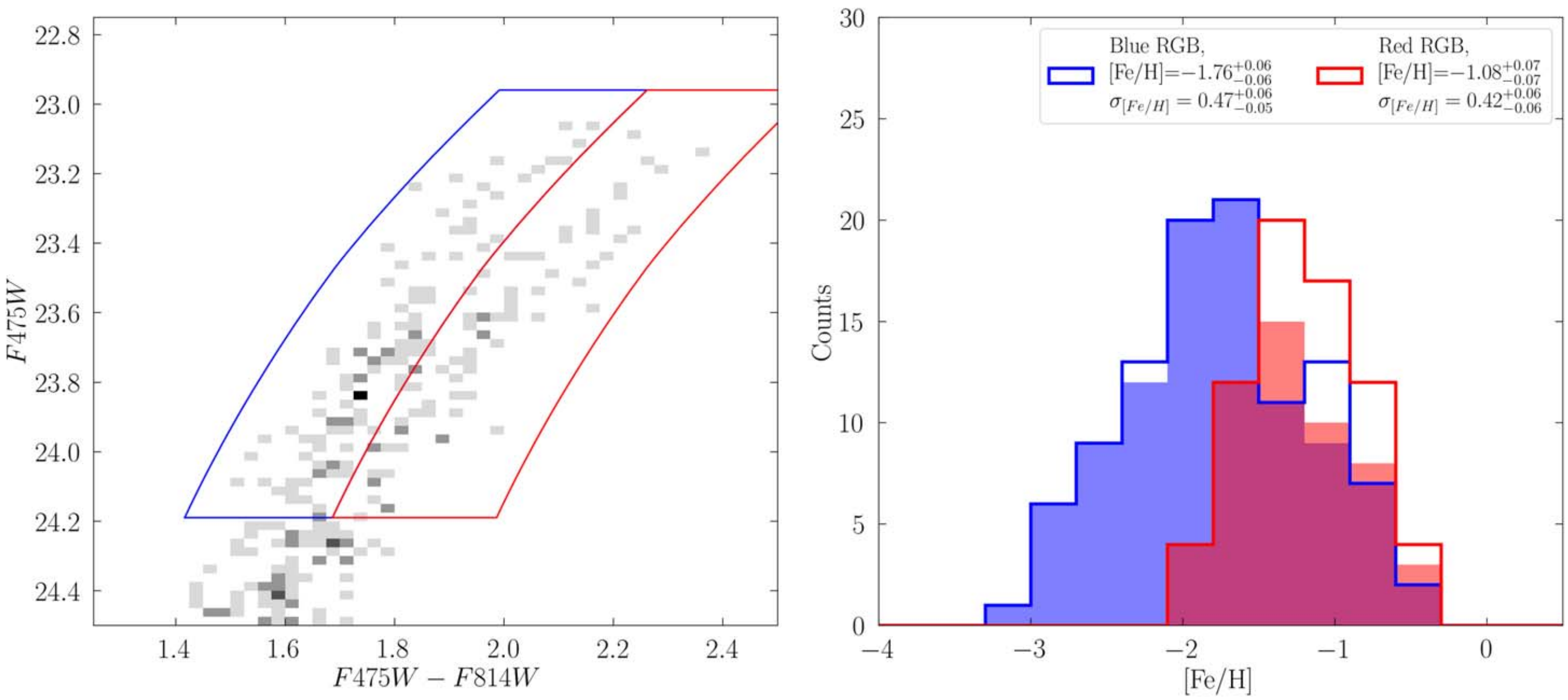

**Figure 9.** MDFs of stars along the respective bifurcations of Tucana's RGB. Left: Hess diagram of Tucana stars analyzed in this work, color-coded by the boxes used to select stars on the blue and red bifurcations of the RGB. We select stars above a certain magnitude range to ensure better completeness for our comparison. Right: comparing the MDFs of stars that fall on the blue and red branches of the bifurcation. Open histograms on the MP and MR ends of the MDF respectively denote upper and lower limits. The metallicity difference between stars on the two branches is apparent even by eye, and the corresponding summary statistics in the legend quantify these differences: the MDF of the blue RGB is more MP than the MDF of the red RGB by ∼0.7 dex.

Additionally, we also investigate the spatial chemical properties of stars in the respective bifurcations and present the results in Figure 10. Even by eye, it is apparent that stars in the MR red RGB are more centrally concentrated than the bluer RGB. This result tracks with our metallicity gradient fit to the entire MDF sample. We also fit the metallicity gradient to stars in the respective bifurcations following the same procedure from Section 4.3. For the blue RGB, we recover $\nabla_{[\mathrm{Fe/H}]} = -0.62 \pm 0.11$ dex $R_e^{-1}$, and for the red RGB, we recover $\nabla_{[\mathrm{Fe/H}]} = -0.42 \pm 0.14$ dex $R_e^{-1}$. We present corresponding correlation plots in Section B.3.

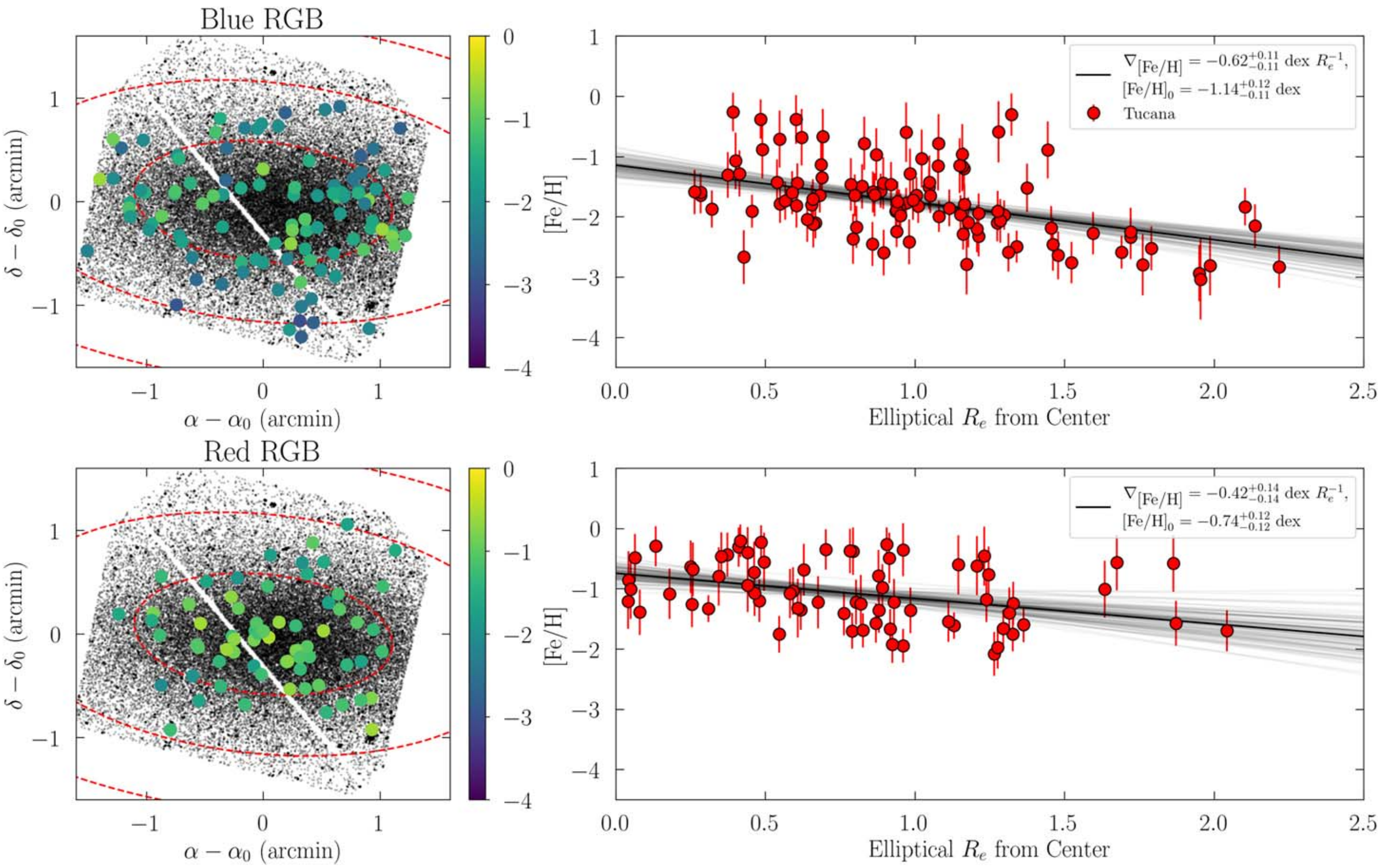

**Figure 10.** Spatial properties of stars along the respective bifurcations of Tucana's RGB. The left panels of each row present the spatial distribution of stars in each bifurcation, color-coded by their metallicity. The right panels of each row present stellar metallicities as a function of elliptical half-light radii and the corresponding metallicity gradient fit. Stars in the more MR red RGB are more centrally concentrated than stars in the blue RGB. We recover a steeper metallicity gradient for stars along the blue RGB than for stars along the red RGB.

## 5. Discussion

### 5.1. Improved MDF Statistics for Tucana and the Power of CaHK

As presented in Section 4.2, we measure for Tucana $\langle[\mathrm{Fe/H}]\rangle = -1.55^{+0.04}_{-0.04}$ and $\sigma_{[\mathrm{Fe/H}]} = 0.54^{+0.03}_{-0.03}$. These measurements are in good overall agreement with the literature values of T20, though with higher precision because we are able to use 374 stars, which is a factor of ∼7 larger than existing CaT spectroscopic measurements. Our mean metallicity measurement is consistent with Tucana's expected location on the dwarf galaxy mass–metallicity relation (Kirby et al. 2013). Our expanded sample size enables us to measure a slight negative skew for our MDF, indicating a longer MP tail for Tucana, but both our skew and kurtosis measurements are consistent with the MDF of Tucana deviating little, if at all, from Gaussianity. In Appendix A, we show how these results largely hold assuming different [α/Fe] versus [Fe/H] relationships.

On the MP end of the MDF, we identify eight EMP star candidates in Tucana, comprising 2% of our sample. This is consistent with observations of other dwarf galaxies in the LG, whose MDFs are generally characterized by a sparse MP tail below $[\mathrm{Fe/H}] < -3.0$ (e.g., Lemasle et al. 2012, 2014; Starkenburg et al. 2013; Hill et al. 2019; Han et al. 2020). EMP stars are scientifically interesting for their ability to trace the chemical enrichment pathways of the first stars (Frebel & Norris 2015), and the low-metallicity dSphs have, due to their relatively simpler SFHs compared to the MW, long been a promising site for EMP searches. We include our EMP candidates in Table 5 as priority targets for spectroscopic follow-up.

On the MR end of the MDF, we identify 76 MR ($[\mathrm{Fe/H}] > -1.0$) stars in Tucana, with 37 of them being lower limit constraints. These measurements are not well constrained because they are low S/N and because the discriminating power of the CaHK narrowband filter diminishes at high metallicity. The shape of the MR tail in a dwarf galaxy's MDF is crucial for modeling its chemical evolution (e.g., Sandford et al. 2022) and is therefore also of interest for detailed follow-up. We provide a table of these stars in Table 6 and invite the community for spectroscopic follow-up in the era of extremely large telescopes.

### 5.2. Radial Trends in Metallicity

#### 5.2.1. Comparison to Literature Gradients

Based on the spatial concentration of different populations of HB stars, Savino et al. (2019) suggested that population gradients should be expected in Tucana. However, T20 was unable to definitively resolve differences in metallicity between the inner parts and the outskirts of the galaxy. They fit their data using an error-weighted least-squares method to infer a metallicity gradient of $\nabla_{[\mathrm{Fe/H}]} = -0.16 \pm 0.09$ dex arcmin$^{-1}$ for Tucana members within $\sim 2.5\, R_e$ of the galaxy. These numbers translate into $-0.13 \pm 0.07$ dex $R_e^{-1}$ and $-0.6 \pm 0.4$ dex kpc$^{-1}$.

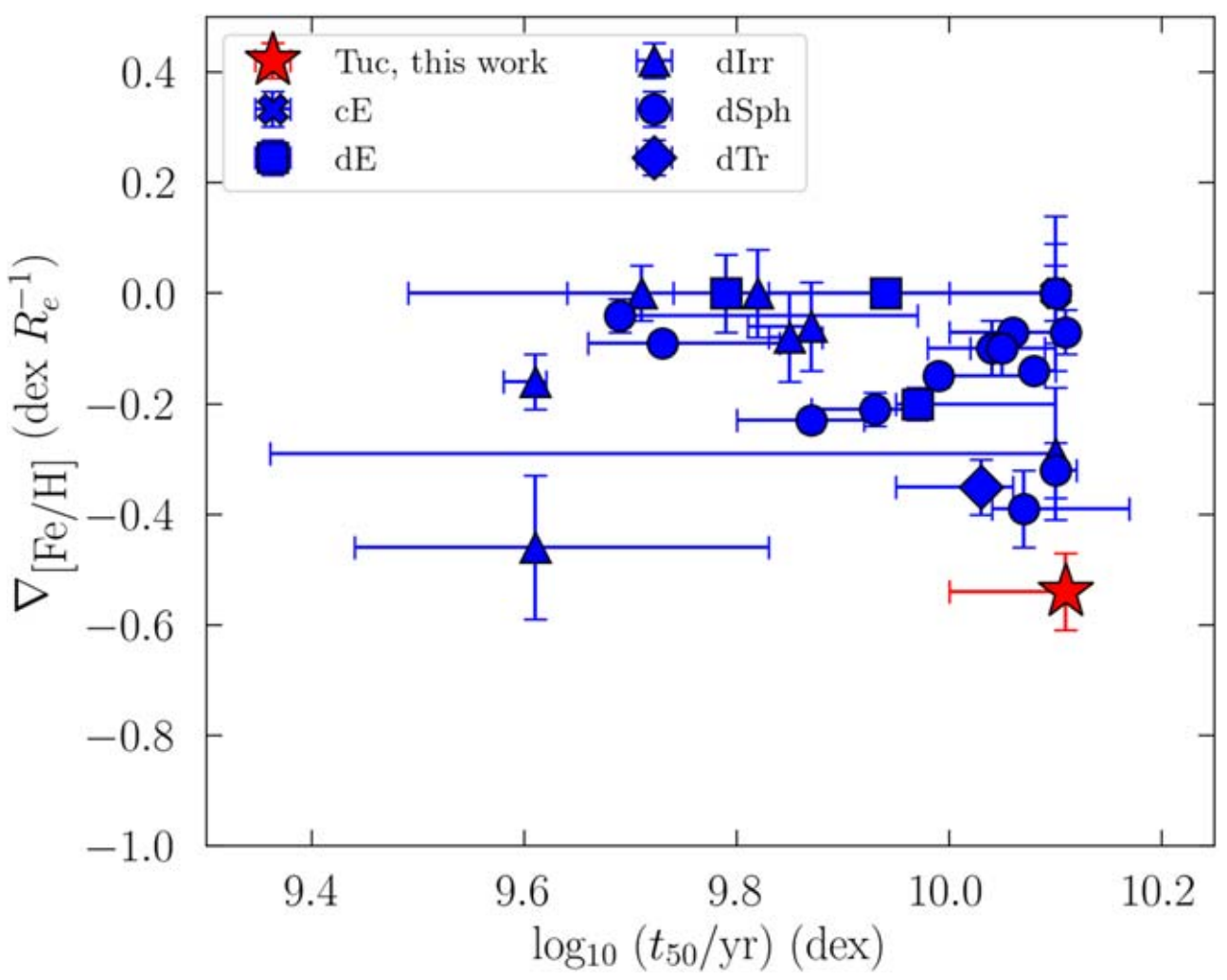

**Figure 11.** Comparing the metallicity gradient in Tucana from our work with gradients of LG dwarf galaxies across a range of morphological type measured with spectroscopy (Taibi et al. 2022). The gradient of Tucana is one of the significant ones known among LG dwarfs. Additionally, while the spectroscopic gradients do not show a clear relation with $t_{50}$ (median stellar age), we demonstrate in Figure 12 that our gradient measurement for Tucana is consistent with predictions from cosmological simulations that posit feedback as the primary mechanism for gradient formation in isolated dwarf galaxies (Mercado et al. 2021).

Due to differences in the treatment of the linear fitting procedure, as well as the galaxy parameters used to convert the gradients into different physical units, we refit a line to the T20 data following the procedures outlined in Section 4.3 to ensure a direct comparison to our measurements and gradient determinations. For the T20 data we find $\nabla_{[\mathrm{Fe/H}]} = -0.10 \pm 0.11$ dex arcmin$^{-1}$, translating to $-0.08 \pm 0.09$ dex $R_e^{-1}$ and $-0.38 \pm 0.44$ dex kpc$^{-1}$, which are slightly different but consistent with the T20 results. Our refit still describes the same qualitative result: the spectroscopic data do not robustly resolve a metallicity gradient in the dwarf, which is not consistent with the gradient we measure from the larger CaHK sample.

Figure 11 compares our metallicity gradient measurements to those from the analysis of archival spectroscopic data carried out by Taibi et al. (2022) in $t_{50}$ versus $\nabla_{[\mathrm{Fe/H}]}$ space for a larger set of LG dwarf galaxies. The $t_{50}$ values for other LG dwarf galaxies come from a combination of the Weisz et al. (2014) compilation, published results from the LCID project (Cetus, Monelli et al. 2010b; Tucana, Monelli et al. 2010b; Leo A, Cole et al. 2007; Hidalgo et al. 2009), and several additional studies (WLM, Albers et al. 2019; Sextans, Bettinelli et al. 2018; And II, Skillman et al. 2017). Individual metallicity measurements used in Taibi et al. (2022) were derived using either spectral fitting of Fe I lines or the CaT calibration. The number of stars used to make metallicity gradient measurements in LG dwarfs ranges from ∼50 to ∼300.

Our metallicity gradient in Tucana is steeper than previous measurements. In fact, it is one of the steepest gradients measured in any LG dwarf galaxy to date. We now consider several aspects of our analysis that could affect the robustness of this result.

First, as shown by Figures 5 and 6, our individual metallicities are consistent with the CaT-based values across the full range of metallicities. Thus, it is unlikely that any systematic in our measurement is the primary driver of a steep gradient.

Second, we assess the impact of selection effects on our measurements. One selection effect impacting our overall analysis is that toward fainter magnitudes, we preferentially lose MR stars in our sample. This is because at a given brightness, MR stars have lower S/N in F395N due to increased absorption in the CaHK features. The effect is similar to the one described in Manning & Cole (2017). To mitigate the impact of this effect in our metallicity gradient inference, we remeasure our metallicity gradient with only stars that are brighter than F475W of 24 mag, where we have higher levels of completeness along the RGB. The resulting gradient is on the order of $-0.8$ dex arcmin$^{-1}$, which is larger in magnitude than the measurement made using the full sample. Thus, it is unlikely that this particular selection effect is inflating our metallicity gradient measurement.

Third, we consider that the discrepancies may result from the uncertain nature of spectroscopic selection effects, which comprise the vast majority of dwarf galaxy metallicity gradient measurements in the literature thus far. In particular, it seems plausible that the empirical picture of gradients has been limited by spectroscopic selection effects, some of them complex (e.g., the placement of stars on masks for multi-object spectroscopy). It becomes difficult to obtain a clear picture of the metallicity gradient when the sampling of stars within particular spatial and magnitude limits is incomplete. In contrast, at a distance of ∼1 Mpc, a single pointing and 12 orbits with HST yield reliable metallicities of RGB stars within $2\, R_e$ of Tucana with a high level of completeness, as illustrated in Figures 3 and 7.

#### 5.2.2. Comparison to Cosmological Simulations

As a means of illustrating the connection of our results to theory, we compare our measured gradient[15] for Tucana to a subset of isolated, simulated dwarf galaxies from the publicly available core suite of FIRE-2 simulations (Wetzel et al. 2023). The simulated galaxies have evolved in isolation and are dispersion-supported systems, which make them good matches to Tucana for this comparison.

Figure 12 compares metallicity gradients from FIRE-2 to our Tucana measurement. In all three panels, we provide the circularly averaged 2D gradient and dispersion measured over 100 random projections for each simulated galaxy. The error bars on the simulations represent the maximum and minimum values. The simulated dwarf galaxies do not display a strong correlation with gradient as a function of stellar mass, and the addition of our Tucana measurement, a single galaxy, does not provide new insight into this trend. The absence of a trend between stellar mass and metallicity has been seen in early empirical results (Taibi et al. 2022) as well as in other simulations (e.g., Schroyen et al. 2013; Revaz & Jablonka 2018).

The center panel compares the Tucana gradient measurement to simulated gradients as a function of median stellar age, $t_{50}$

[15] FIRE-2 dwarf galaxies are known to struggle with producing mean metallicities that are lower than many observed dwarf galaxies. This tension also exists between Tucana and the comparison simulations. Since we are only interested in metrics of relative metallicity across the galaxy, we do not include mean metallicity comparisons in our discussion. Efforts to resolve agreement in mean metallicities are well underway, e.g., Gandhi et al. (2022).

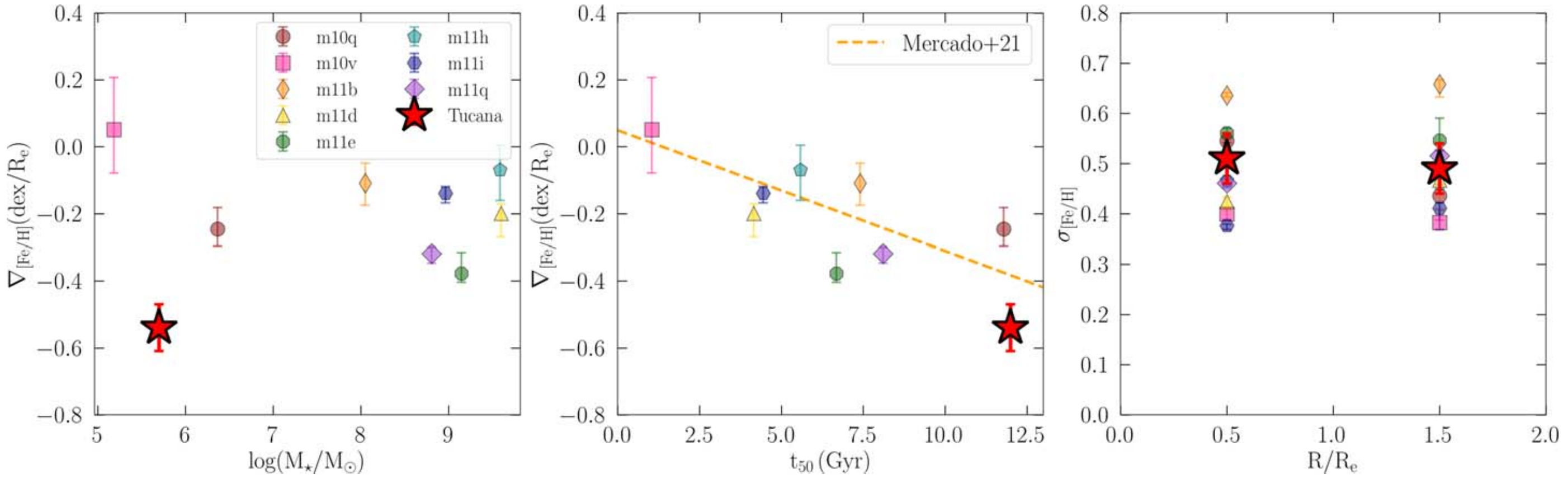

**Figure 12.** Comparing our metallicity gradient measurement for Tucana with those from the FIRE-2 simulations (e.g., Mercado et al. 2021), which posit that metallicity gradients in isolated dwarf galaxies are driven by feedback-induced breathing modes (El-Badry et al. 2016). Left: comparing our gradient measurement and gradients from simulations as a function of stellar mass. The weak correlation between stellar mass and gradient strength is an emerging consensus among simulations, and we present this panel largely for completeness reasons. Center: comparing our gradient measurement and gradients from simulations as a function of $t_{50}$ (Gyr). The orange dotted line is the relation between $t_{50}$ and gradient strength as found by Mercado et al. (2021). Our gradient measurement for Tucana is consistent with this theoretical relation. Right: examining trends in metallicity dispersion as a function of $R_e$. The radial dispersion trend in Tucana is consistent with those from simulations.

(Monelli et al. 2010a). In the FIRE-2 simulations, these two parameters are strongly correlated, where galaxies with older stellar populations have stronger gradients. This relationship was first pointed out by Mercado et al. (2021) using a similar sample of FIRE-2 simulated dwarf galaxies not included in the public data release. Because star formation tends to flatten preexisting radial trends, metallicity gradients are best preserved in the oldest galaxies that stopped forming stars early on. We include the trend that they report in this panel as the orange dashed line and find that the simulations we use are in good agreement with their study. The T20 gradient for Tucana ($-0.13 \pm 0.07$ dex $R_e^{-1}$) is too shallow to be in agreement with this relationship, whereas our measurement for Tucana's gradient affirms it well.

Our well-sampled MDFs also enable measurements of metallicity dispersion as a function of radial extent. In the right panel, we make comparisons between our measurements and those from the FIRE-2 simulations. Our results are comparable to those of the simulations, where the gradients within the inner $R_e$ of a galaxy and the gradient within the adjacent bin are comparable.

In FIRE-2 simulated dwarf galaxies, metallicity gradients are formed primarily through feedback from successive episodes of star formation, which add energy to the orbits of older (i.e., more MP) stars and drive outward radial migration over the course of a galaxy's lifetime (e.g., El-Badry et al. 2016). Metallicity gradient measurements made from a complete sample of stars further down the LF of isolated LG dwarf galaxies can better determine whether this prediction bears out in the dwarf galaxy population more generally and not just in the case of Tucana.

Radial metallicity dispersion trends have not been previously well quantified in a dwarf galaxy due to the limited sample sizes observed. Similarly, radial dispersion trends have also not been examined in simulations prior to this work. Given that not all simulations agree that feedback drives gradient formation (e.g., Marcolini et al. 2008; Sales et al. 2010; Hausammann et al. 2019; Munshi et al. 2021), metallicity dispersions are another metric that may be useful for understanding the interplay of different processes that drive or wash out dwarf metallicity gradients. Our results for Tucana reveal the promise of CaHK narrowband imaging for enabling the exploration of radial metallicity dispersion trends in dwarf galaxies across the LG, and we recommend further exploration of this metric as part of developing the picture of observational imprints from different processes that drive spatial trends within a dwarf galaxy.

### *5.3. Metallicity Gradients as Probes of Cored Dark Matter Profiles*

The "too big to fail" (TBTF) problem (Boylan-Kolchin et al. 2011) describes the overdensity of dark matter (DM) halos in simulated dwarf galaxies compared to the actual observed central DM density of dwarf galaxies based on velocity measurements. Efforts to reconcile theory to observation have invoked the chaotic effects of baryonic processes, particularly whether the energetics of SNe can sufficiently reduce the central DM density of halos to form cores, though the extent to which this is possible remains a point of controversy (Garrison-Kimmel et al. 2013; Madau et al. 2014).

The most direct probe of central DM density is stellar velocity measurements, and in the case of Tucana, the central density of its DM halo has been a subject of controversy within spectroscopic studies: it is either an exception to the TBTF problem compared to its other LG analogs because it has a dense central DM halo (Gregory et al. 2019), or its properties are consistent with the properties of LG dwarfs (T20). These studies have been limited by sample size because of the limited ability for spectroscopy to reach an adequate number of stars to construct a high-confidence, detailed profile.

Given the current limitations of spectroscopic studies in this arena, it is important to make predictions for ancillary observations that could weigh in on this theoretical picture. Thus, an important contribution of El-Badry et al. (2016) is proposing that age and metallicity gradients are also an observational consequence of cored DM halo formation from SN feedback. Age gradients have historically been difficult to detect in Tucana due to degenerate effects in age and metallicity, particularly for old stellar populations (Hidalgo et al. 2013), and evidence of a metallicity gradient was detected using the different concentrations of HB stars of different metallicity and age, but it was not well quantified (Harbeck et al. 2001; Savino et al. 2019).

As established previously, we robustly detect a metallicity gradient in Tucana. The gradient is consistent with predictions from the FIRE simulations, which produce both gradients and DM cores through the same baryonic processes. With the advent of high-confidence dwarf metallicity gradients measured using CaHK imaging, we propose additional theoretical exploration into the utility of metallicity gradients as an indirect probe for the underlying kinematics of dwarf galaxy DM halos. Any such theory will also need to account for the interpretation of metallicity gradients and the different kinematic properties of MP and MR components that are being robustly detected in nearby MW satellites (Pace et al. 2020; Tolstoy et al. 2023).

### 5.4. The MDF of Tucana in Context

Tucana's isolated, quenched status makes it an outlier compared to its field star-forming analogs, as well as its satellite quenched analogs. Its contested status as a backsplash galaxy (Teyssier et al. 2012; Santos-Santos et al. 2023), as opposed to a galaxy that has truly evolved in the field away from the influence of more massive hosts, also complicates its interpretation. Here, we compare the broad features of its MDF to those of other LG dwarf galaxies to determine the potential insights offered.

First, we compare the MDF of Tucana to the MDFs of MW dSphs that are close to mass analogs of Tucana. We reference the detailed MDFs measured by Kirby et al. (2013). Whereas the MDF of Tucana does not deviate strongly from Gaussianity, the MDF of Draco demonstrates a long MP tail. The MDF of Ursa Minor is closer to symmetrical, though it has a slightly stronger MR tail. Although their SFHs are comparable to Tucana's in duration and age, the mean metallicities of these galaxies are lower than Tucana's by ∼0.4 dex, suggesting that they experienced slower chemical enrichment over the course of their lifetimes compared to their isolated counterpart. This may be related to different mechanisms driving the depletion of gas in these dwarf galaxies: while Draco and UMi could have had their gas stripped by the proto-MW, Tucana may have consumed gas largely through star formation, resulting in a higher overall metallicity for its stellar population.

Second, we compare the MDF of Tucana to the MDFs of other isolated dwarf galaxies on the periphery of the LG. These galaxies are all currently forming stars. Kirby et al. (2017) present spectroscopic metallicities of Leo A, Aquarius, and the SagDIR, all of which have stellar masses comparable to Tucana but are currently star-forming. The resulting MDFs of 113, 23, and 43 stars, respectively, appear qualitatively similar to that of Tucana's, in that they do not display skewness. Tucana also has a mean metallicity comparable to that of Leo A and Aquarius to within ∼0.1 dex and is more MR than SagDIG by 0.3 dex.

The similar metallicity properties between Tucana and some of its star-forming analogs is interesting to consider further. Taken at face value, the implication is that field dwarf galaxies of different morphological properties end up at comparable metallicities, whether they arrived there 10 Gyr ago after 3–4 Gyr of star formation (e.g., Tucana; Monelli et al. 2010a) in isolation, or whether they have been slowly forming stars throughout cosmic time and recently accelerated the process within the last 3 Gyr (e.g., Leo A; Cole et al. 2007). This comparison implies that there are enrichment processes operating on different timescales that can produce degenerate observational signatures in [Fe/H].

Articulated in a different way, the evolutionary pathways of star-forming isolated galaxies in the near-present is presumably different from the evolutionary pathways of isolated galaxies quenched in the distant past. In this framing, the interpretation of Tucana as a splashback galaxy that was previously bound to M33 (e.g., Teyssier et al. 2012; Santos-Santos et al. 2023) can rule out evolutionary narratives contingent upon isolation, but it alone would not be able to provide a full detailed account of Tucana's chemical enrichment in comparison to its LG stellar mass analogs. Addressing these questions would also have implications for our detailed understanding of the physics setting the present-day mass–metallicity relation for dwarf galaxies, which generally holds across a range of morphological types (Kirby et al. 2013).

### 5.5. Interpretation of the Split RGB

In this section, we discuss the possible evolutionary histories of Tucana implied by observations of its split RGB. We show in Section 4.4 that the bifurcation is driven in large part by metallicity differences between the two respective populations. Additionally, stars belonging to the MP blue branch are more spatially extended than stars in the red branch (Figure 10), affirming our measurement of a significant global gradient in Tucana. Given that chemodynamical populations have been found in other dwarf galaxies (e.g., Pace et al. 2020; Tolstoy et al. 2023), we compute velocity and velocity dispersions for stars in the respective bifurcations using data from T20. We find no kinematic difference between these populations, although this calculation is likely limited due to the small number of stars with velocities in Tucana.

We note that the MDF of the blue RGB (Figure 9) has a comparable mean metallicity and dispersion to the MDFs of the MW satellites. That is, removal of the red RGB from Tucana would produce a stellar population whose global MDF properties more closely resemble Ursa Minor or Draco. As the ratio of blue and red RGB stars are roughly of order unity in our sample, we proceed assuming that their populations are roughly $10^5\,M_\odot$ each in stellar mass.

We consider how these features could have formed in situ in Tucana. The SFH of Tucana from Savino et al. (2019) uses HB stars to effectively resolve two major star-forming episodes that are spaced roughly 1–2 Gyr apart. Following this finding, we consider the possibility that each respective bifurcation corresponds to one star formation episode, with the blue MP RGB corresponding to the earlier episode and the red MR RGB corresponding to the more recent episode. The first round of star formation produces a more MP MDF resembling those of MW satellites. During the quiescent period, the ISM is enriched further by a number of SNe, including Type Ia, so that the second episode of star formation proceeds from a more MR ISM and results in the MDF traced out by the red RGB. Qualitatively, this picture would also be consistent with the formation of feedback-driven metallicity gradients in dwarf galaxies, where older, MP stars are more spatially extended than younger, MR stars. Chemical evolution modeling that treats each component separately should ascertain the feasibility of this scenario and possibly characterize the physics driving star formation at different epochs of Tucana's lifetime.

We next consider the possibility that these two RGB branches are the result of the merging of two distinct populations. Simulation literature is not conclusive about the ubiquity of low-mass dwarf galaxy mergers, with some studies suggesting that dwarf galaxy formation is characterized by numerous merger events (e.g., Benítez-Llambay et al. 2016; Jeon et al. 2017) and others strongly discouraging this scenario (e.g., Fitts et al. 2018). Given this ambiguity, we consider this possibility for narrative completeness. At a stellar mass of roughly $10^5\,M_\odot$, the blue RGB of Tucana could have the properties of a dwarf galaxy consistent with the mass–metallicity relation. On the other hand, the red RGB would be too MR compared to expectations for a dwarf galaxy (i.e., from the mass–metallicity relation), and its large metallicity dispersion rules out the possibility that it is an MR star cluster. Thus, a theoretical stellar population represented by the red RGB would be an outlier in the current observational and theoretical landscape of galaxy formation. On the other hand, the blue RGB has a steeper metallicity gradient than UMi (Taibi et al. 2022, using Pace et al. 2020 spectroscopic data subtending $\sim 2.5\,R_e$ of the dwarf), so the blue RGB is not a perfect analog to known dSphs. The sum of this evidence suggests that it is unlikely that Tucana's bifurcated RGB is the product of low-mass galaxy mergers.

More broadly, we situate the split RGB feature within the broader landscape of known dwarf galaxy stellar populations. The split RGB feature is itself rare, and to date, other galaxies that display this feature are the Sextans dwarf spheroidal (Bellazzini et al. 2001) and the And II dwarf galaxy (McConnachie et al. 2007; Skillman et al. 2017). On the other hand, morphologically distinct features that trace out separate stellar populations are common in LG dwarf galaxies, even if they do not manifest in the RGB. For example, the Carina dSph, with multiple discrete bursts of SFH and corresponding MSTO sequences, does not display a split in its RGB (de Boer et al. 2014). Why some galaxies have split RGBs while others, with extended SFHs, metallicity dispersions, and other distinct morphological features in their CMDs, do not is still an outstanding question. While we have demonstrated with Tucana that the level of observed separation is driven in large part by metallicity differences of $\sim 0.7$ dex between the two populations, it is possible that a combination of higher-order effects such as age and $\alpha$ may influence the prevalence of split RGBs more broadly. These effects are not yet well understood and require additional studies.

## 6. Conclusion

In this paper, we present the MDF of the isolated quenched dwarf galaxy Tucana measured using HST CaHK narrowband imaging. In a single HST pointing, we are able to measure metallicities for stars brighter than 25 mag in F475W within $\sim 2.5\,R_e$ of Tucana to a higher level of completeness than has been previously possible via spectroscopic surveys. From this data set with vastly improved sampling both along the luminosity function of the galaxy as well as its spatial extent, we present the key conclusions for Tucana.

1. From a sample of 374 stars, we measure $\langle [\mathrm{Fe/H}] \rangle = -1.55^{+0.04}_{-0.04}$ and $\sigma_{[\mathrm{Fe/H}]} = 0.54^{+0.04}_{-0.04}$. Additionally, we quantify higher-order moments in the MDF such as skew and kurtosis and find that the MDF of Tucana does not deviate significantly from a Gaussian.
2. Across the $\sim 2.5\,R_e$ subtended by our imaging, we measure $\nabla_{[\mathrm{Fe/H}]} = -0.54 \pm 0.07$ dex $R_e^{-1}$. In circular units, we detect $\nabla_{[\mathrm{Fe/H}]} = -0.57 \pm 0.1$ dex arcmin$^{-1}$, which corresponds to $\nabla_{[\mathrm{Fe/H}]} = -2.1 \pm 0.3$ dex kpc$^{-1}$ in units of physical distance.
3. We characterize the metallicity features of stars along the respective bifurcations of Tucana's RGB. The blue RGB has $\langle [\mathrm{Fe/H}] \rangle = -1.78^{+0.07}_{-0.06}$ and $\sigma_{[\mathrm{Fe/H}]} = 0.44^{+0.07}_{-0.06}$, whereas the red RGB has $\langle [\mathrm{Fe/H}] \rangle = -1.08^{+0.06}_{-0.06}$ and $\sigma_{[\mathrm{Fe/H}]} = 0.42^{+0.06}_{-0.06}$. Stars in the more MP blue RGB are also more spatially diffuse than stars in the red RGB, consistent with our metallicity gradient fit to the global sample. For the blue and red RGB stars, we measure $\nabla_{[\mathrm{Fe/H}]} = -0.62 \pm 0.11$ dex $R_e^{-1}$ and $\nabla_{[\mathrm{Fe/H}]} = -0.42 \pm 0.14$ dex $R_e^{-1}$, respectively.
4. We compare our gradient measurement of Tucana to simulated gradients of isolated dwarf galaxies from the FIRE-2 suite and find that our measurement affirms the gradient strength–median stellar age law as found by Mercado et al. (2021). As the same feedback mechanisms driving gradients are also expected to form cored DM halo profiles that might alleviate small-scale tensions with ΛCDM (El-Badry et al. 2016; Bullock & Boylan-Kolchin 2017), we recommend additional studies investigating possible relationships between metallicity gradient strength and host halo properties.
5. Compared to the MDF of MW dSphs Ursa Minor and Draco, most similar in mass and SFH to Tucana, the MDF of Tucana is more MR on average by $\sim 0.4$ dex and more consistent with Gaussianity, implying that MW satellite dwarfs may have enriched more slowly than field dwarfs throughout a similar period of star formation. On the other hand, removal of the red RGB component in Tucana would result in an MDF more consistent with those of MW satellites.
6. We compare the MDF of Tucana to the MDF of other star-forming LG dwarf galaxies of similar mass (Kirby et al. 2017) and find qualitatively similar results, suggesting that the physics driving different evolutionary pathways of field dwarf galaxies across morphological type can produce degenerate [Fe/H] signatures observed in the present day.
7. Chemical evolution modeling of the MDF of Tucana will be essential for obtaining a full detailed picture of the physics driving Tucana's observed properties. Constraints on the SN energetics governing Tucana's chemical evolution may also inform whether such processes can give rise to the observed metallicity gradient. Modeling efforts will also have to account for the split RGB and the attendant metallicity properties of each bifurcation. These efforts will have strong scientific synergy with forthcoming chemical abundance measurements, particularly $[\alpha/\mathrm{Fe}]$, from JWST GO-3788 (PI: Weisz), as well as future efforts to push stellar spectroscopy to further depths within the LG and beyond (Sandford et al. 2020).

More broadly, our work affirms the power of CaHK narrowband imaging to measure stellar metallicities and recover spatially resolved stellar MDFs of distant galaxies at a higher efficiency compared to what has been previously done. This technique will continue to be highly complimentary with science capacities from the newly launched JWST for mapping the detailed chemistry of distant dwarf galaxies of high scientific value. Homogeneous metallicity and chemical abundance measurements and attendant chemical evolution studies of LG dwarf galaxies will be essential for fleshing out the full explanatory picture of the observed diversity of dwarf galaxies across mass, morphological type, and environment. Their application to interpreting observations of distant dwarf galaxies will realize the promise and power of near-field cosmological studies for providing detailed insight into the evolution of dwarf galaxies across cosmic time (e.g., Boylan-Kolchin et al. 2016).

## Acknowledgments

We thank the anonymous referee whose comments enhanced the readability of this manuscript. We also thank Evan Skillman for additional comments that improved the clarity of this draft. S.W.F. acknowledges support from a Paul & Daisy Soros Fellowship and from the NSFGRFP under grants DGE 1752814 and DGE 2146752. S.W.F., D.R.W., M.B.K., A.S., and N.R.S. acknowledge support from HST-GO-15901, HST-GO-16226, and HST-GO-16159 from the Space Telescope Science Institute, which is operated by AURA, Inc., under NASA contract NAS526555. E.S. acknowledges funding through VIDI grant "Pushing Galactic Archaeology to its limits" (with project No. VI.Vidi.193.093), which is funded by the Dutch Research Council (NWO).

This work made use of the Savio computational cluster provided by the Berkeley Research Computing Program at the University of California, Berkeley. Some/all of the data presented in this paper were obtained from the Mikulski Archive for Space Telescopes (MAST) at the Space Telescope Science Institute. The specific observations analyzed can be accessed via the following link: 10.17909/8974-w227. STScI is operated by the Association of Universities for Research in Astronomy, Inc., under NASA contract NAS526555. Support to MAST for these data is provided by the NASA Office of Space Science via grant NAG57584 and by other grants and contracts.

*Software:* DOLPHOT (Dolphin 2000, 2016), astropy (Astropy Collaboration et al. 2013; The Astropy Collaboration et al 2018; Astropy Collaboration et al. 2022), numpy (Oliphant 2006), matplotlib (Hunter 2007), emcee (Foreman-Mackey et al. 2013), corner (Foreman-Mackey 2016), scipy (Virtanen et al. 2020).

## Appendix A
## Impact of $\alpha$ Assumptions on MDF Measurement

In this section, we present MDF measurements of Tucana from assuming different levels of [$\alpha$/Fe] across all of its stars. Figure 13 presents the results of MDF inference against different [$\alpha$/Fe] relationships, and Table 3 tabulates the corresponding MDF summary statistics. The upper row is our fiducial assumption, while the middle and bottom rows shift the location of the ankle and knee by 0.5 dex backward and forward, respectively. The changes in mean metallicity and metallicity dispersions are attributed to the recovery of MR stars that determine the strength of the MR tails: an MP knee results in the recovery of more MR stars, while an MR knee results in the recovery of fewer MR stars. In any case, the corresponding mean metallicity and metallicity dispersions are still within $2\sigma$ agreement with the values of the MDF measured from the fiducial [$\alpha$/Fe] assumption. The skew and kurtosis are also altered between different assumptions but remain consistent with one another to $2\sigma$. We find that different assumptions for the [$\alpha$/Fe] versus [Fe/H] relationship do not fundamentally change the conclusions presented in this paper.

**Table 3**
Tucana MDF Assuming Different [$\alpha$/Fe] vs. [Fe/H] Relationships

| Feature | Parameter | Value |
|---|---|---|
| Fiducial | $\langle$[Fe/H]$\rangle$ | $-1.55^{+0.04}_{-0.04}$ |
| | $\sigma_{[Fe/H]}$ | $0.54^{+0.03}_{-0.03}$ |
| | Skew | $-0.16^{+0.11}_{-0.11}$ |
| | Kurtosis | $-0.02^{+0.28}_{-0.21}$ |
| [$\alpha$/Fe] knee, ankle = −2.5, −1.5 | $\langle$[Fe/H]$\rangle$ | $-1.46^{+0.04}_{-0.04}$ |
| | $\sigma_{[Fe/H]}$ | $0.61^{+0.04}_{-0.04}$ |
| | Skew | $0.0^{+0.1}_{-0.1}$ |
| | Kurtosis | $-0.1^{+0.2}_{-0.2}$ |
| [$\alpha$/Fe] knee, ankle = −1.5, −0.5 | $\langle$[Fe/H]$\rangle$ | $-1.59^{+0.03}_{-0.03}$ |
| | $\sigma_{[Fe/H]}$ | $0.47^{+0.03}_{-0.03}$ |
| | Skew | $-0.2^{+0.1}_{-0.1}$ |
| | Kurtosis | $0.1^{+0.3}_{-0.3}$ |

**Note.** Summary of metallicity properties of Tucana's MDF, assuming different [$\alpha$/Fe] to [Fe/H] relations.

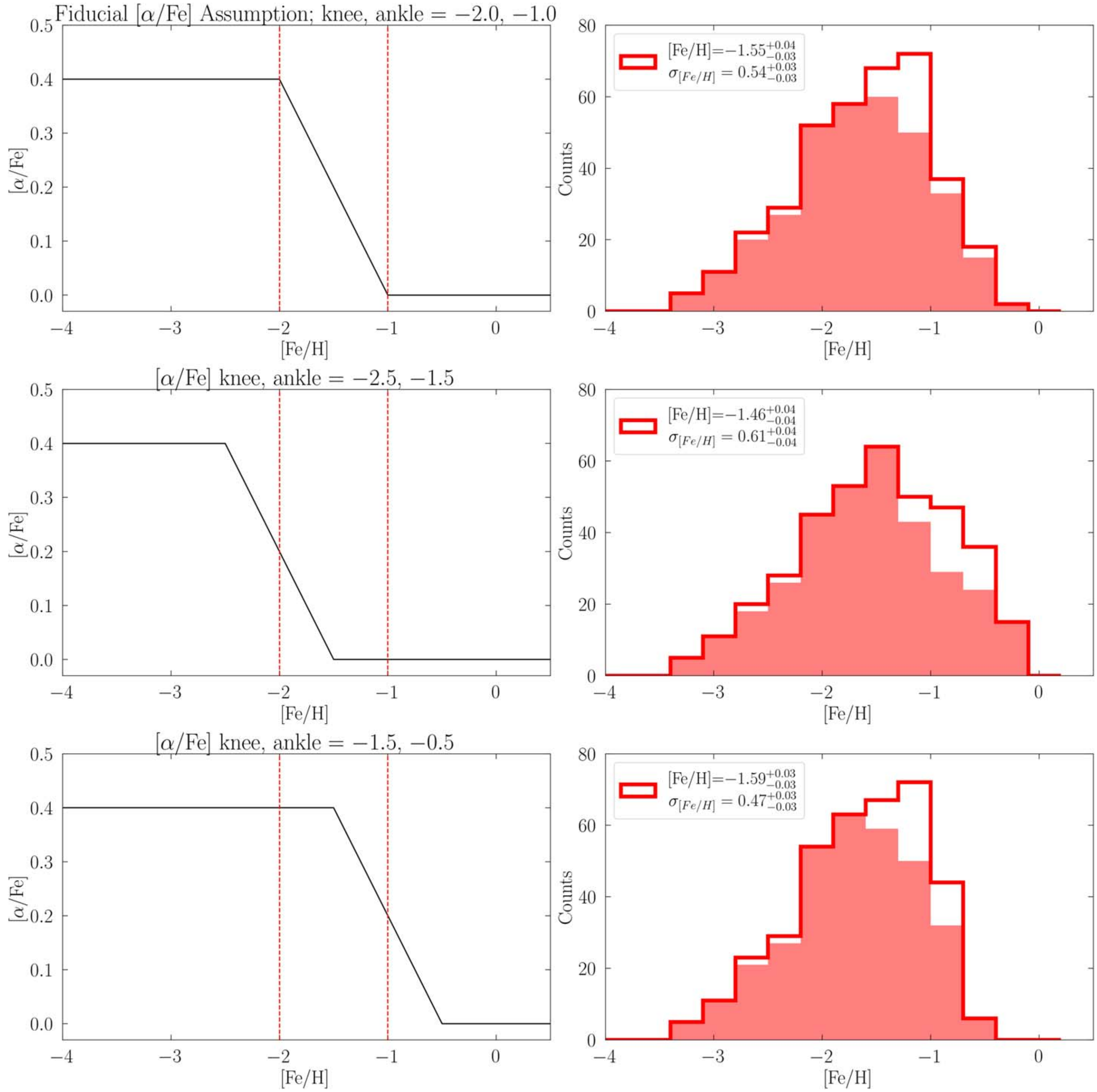

**Figure 13.** Impact of varying the [$\alpha$/Fe] assumption on the recovered MDF of Tucana. The top row presents our fiducial assumption, while the middle and bottom rows present the recovered mean metallicity and dispersion by assuming a relation that shifts the ankle and knee by 0.5 dex backward and forward, respectively. Additional summary statistics are presented in Table 3. The different $\alpha$ do not change the major conclusions of this work.

# Appendix B
# Metallicity Gradient Measurements

## B.1. Metallicity Gradient Corner Plots

Figure 14 presents the corner plot corresponding to the metallicity gradient value reported in the main body of this paper. The figure shows correlations between the intercept, gradient, and a fractional uncertainty underestimation value inferred by the gradient measurement procedure outlined in Section 4.3.

## B.2. Metallicity Gradient Using Circular Arcminutes

We provide fits to the metallicity gradient in units of on-sky distance, or arcminutes, from the center of Tucana, following the method outlined in Section 4.3. By nature of this metric, this gradient fit necessarily assumes circular distance. Figure 15 shows the resulting gradient fit against the data, and Figure 16 shows the corresponding corner plots. The fit that we find in this case is comparable to the fit using elliptical half-light radii.

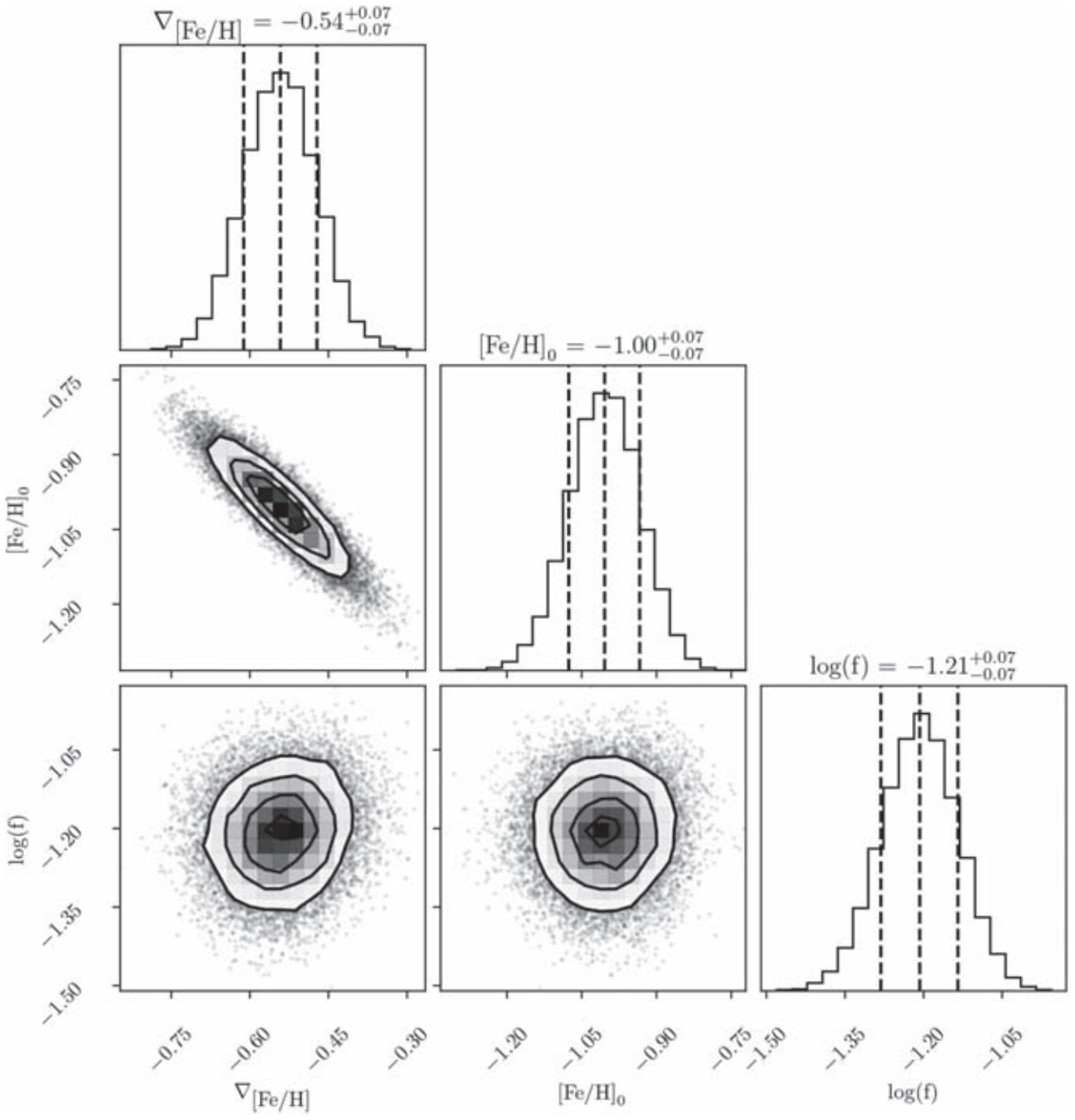

**Figure 14.** Corner plot of metallicity gradient fit using elliptical $R_e$.

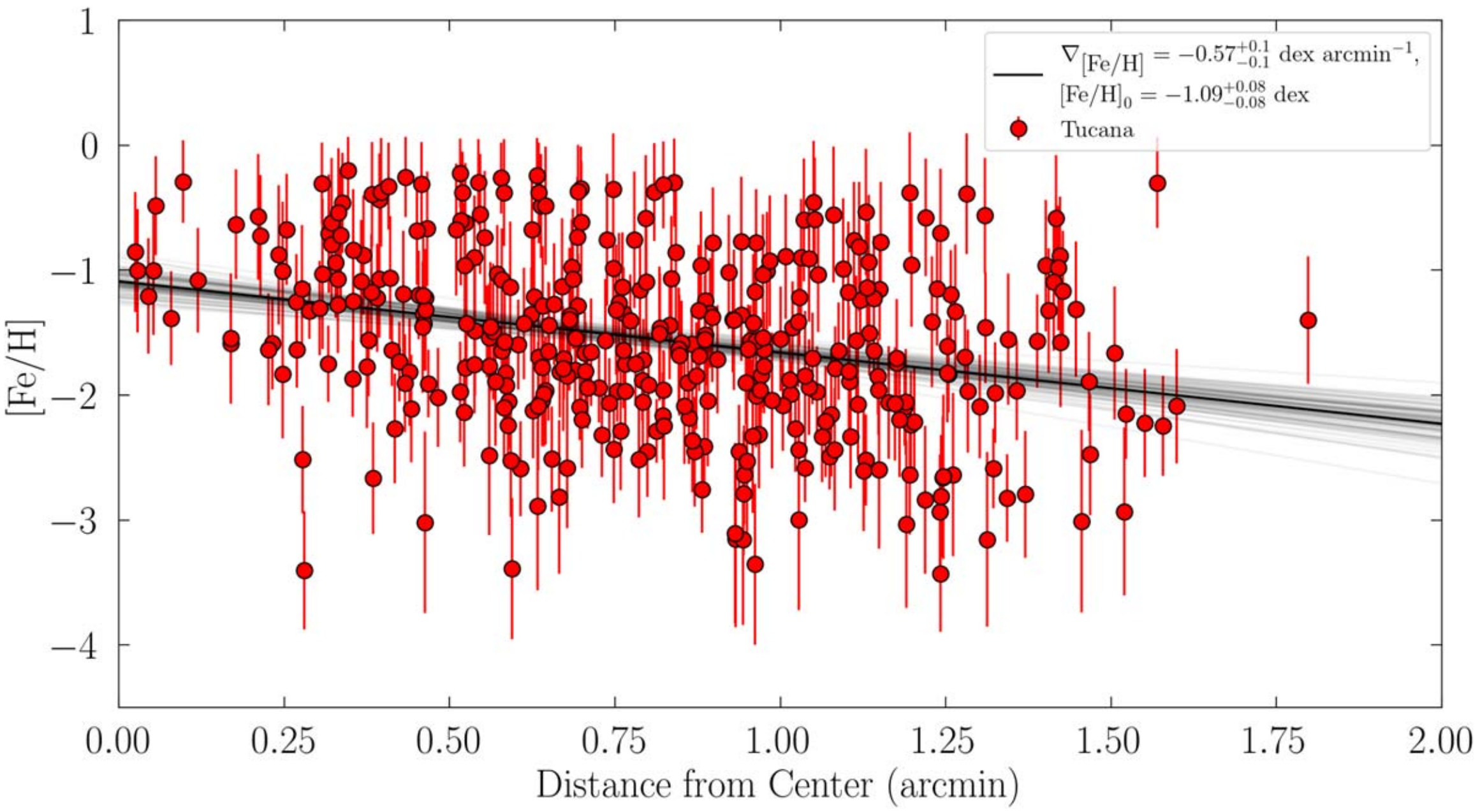

**Figure 15.** Result of fitting line to our Tucana data as a function of circular arcminutes from the center. The gradient from this method is comparable to the gradient from fitting as a function of elliptical half-light radius.

### B.3. Metallicity Gradient Corner Plots for Split RGB Components

In Figures 17 and 18, we present the correlation plots corresponding to the metallicity gradient fits to Tucana's split RGB components.

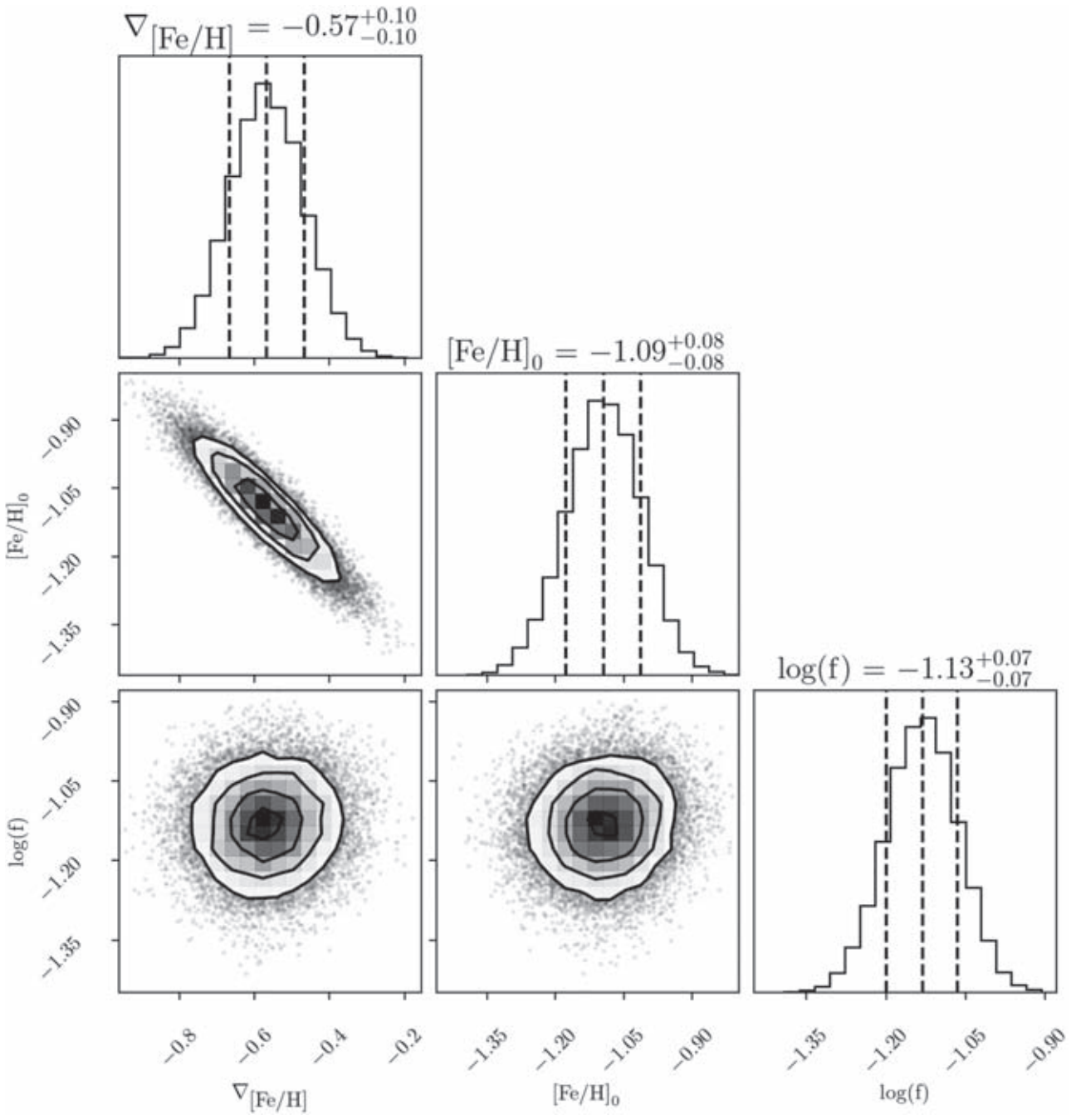

**Figure 16.** Corner plot of metallicity gradient fit using distance from the center of Tucana in circular arcminutes.

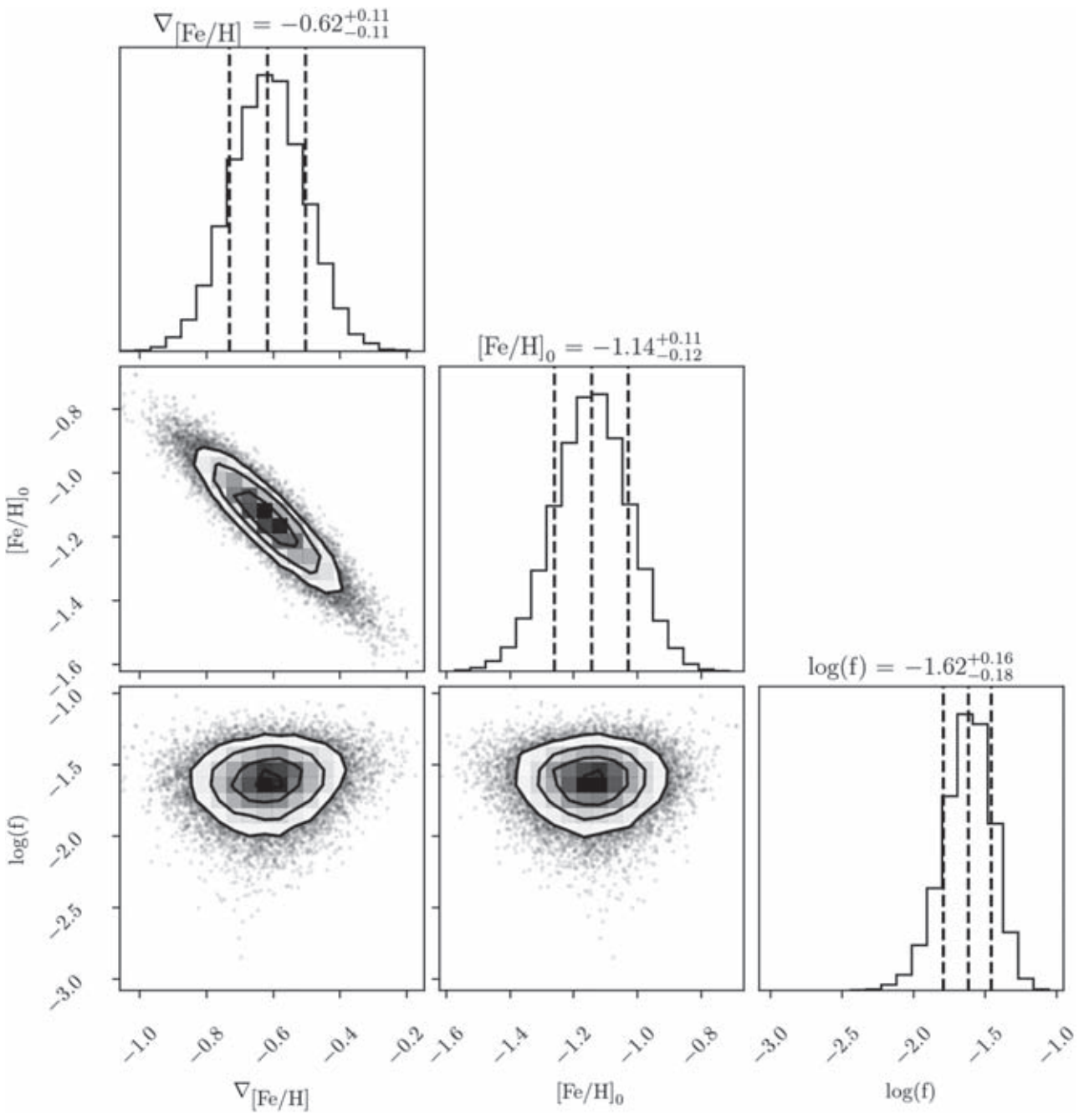

**Figure 17.** Corner plot of metallicity gradient fit to Tucana's blue RGB, using elliptical $R_e$ distance from the center of Tucana.

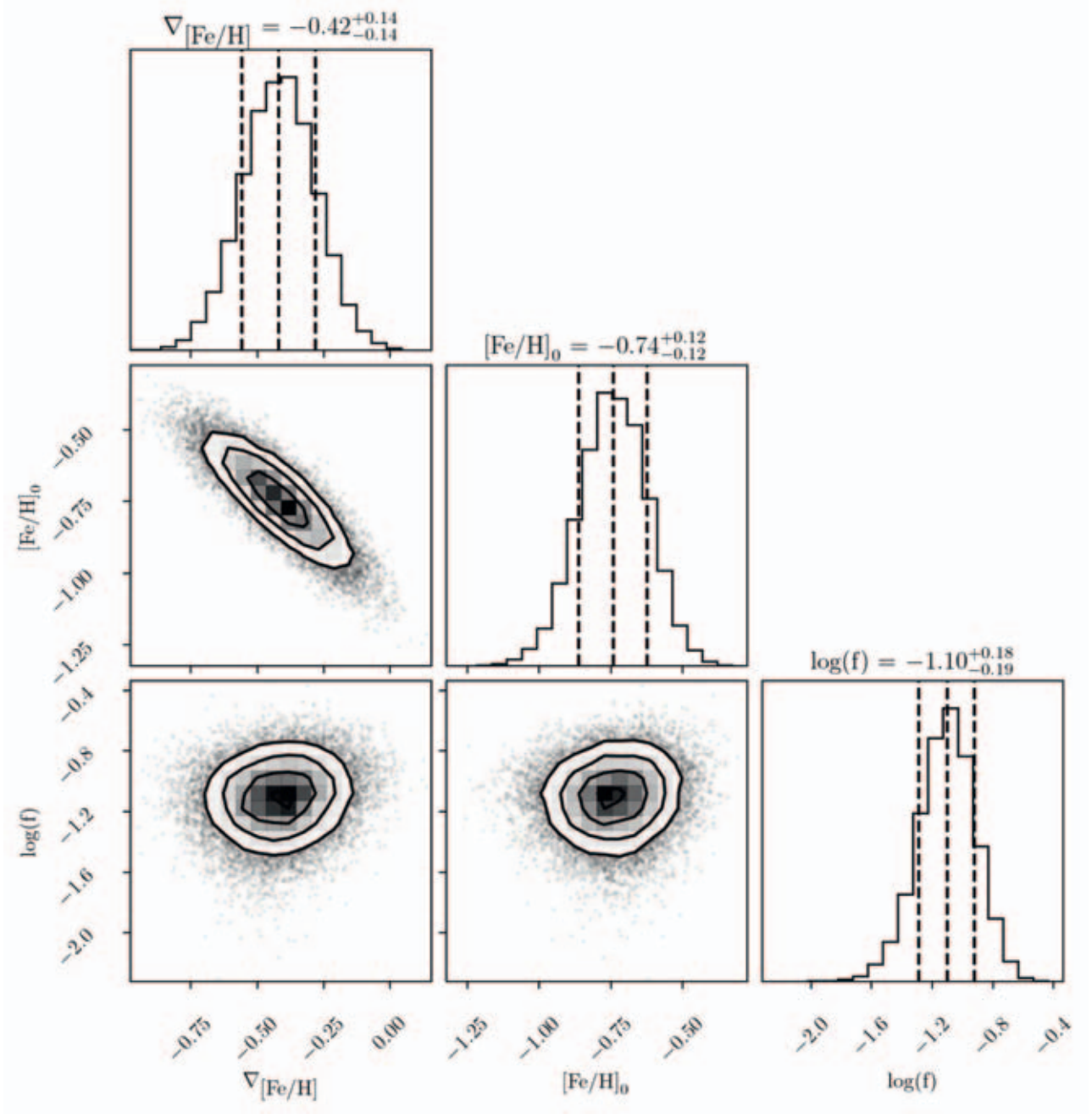

**Figure 18.** Corner plot of metallicity gradient fit to Tucana's red RGB, using elliptical $R_e$ distance from the center of Tucana.

## Appendix C
## Table of Metallicity Measurements

In this section, we present the tables of individual stellar metallicity measurements. Table 4 presents measurements for all the stars in our sample, Table 5 presents the EMP star candidates identified in this work, and Table 6 presents the MR stars in Tucana.

**Table 4**
Metallicity Measurements for All Stars

| Star | R.A. (deg) | Decl. (deg) | F814W (mag) | F475W (mag) | F395N (mag) | VI (mag) | CaHK (mag) | [Fe/H] (dex) | Split RGB? |
|---|---|---|---|---|---|---|---|---|---|
| 0 | 340.467365 | −64.418107 | 20.760 ± 0.001 | 23.132 ± 0.002 | 25.762 ± 0.110 | 2.372 ± 0.002 | −0.928 ± 0.110 | $-1.33^{+0.12}_{-0.09}$ ± 0.2 (syst.) | R |
| 1 | 340.503553 | −64.424886 | 20.911 ± 0.001 | 23.068 ± 0.003 | 25.397 ± 0.129 | 2.157 ± 0.003 | −0.907 ± 0.129 | $-1.19^{+0.33}_{-0.26}$ ± 0.2 (syst.) | B |
| 2 | 340.446929 | −64.422353 | 20.876 ± 0.001 | 23.150 ± 0.002 | 26.046 ± 0.131 | 2.274 ± 0.002 | −0.515 ± 0.131 | $>-0.83$ | R |
| 3 | 340.408890 | −64.416230 | 20.848 ± 0.001 | 23.075 ± 0.002 | 25.281 ± 0.065 | 2.227 ± 0.002 | −1.134 ± 0.065 | $-1.61^{+0.11}_{-0.06}$ ± 0.2 (syst.) | R |
| 4 | 340.442112 | −64.423851 | 20.899 ± 0.001 | 23.176 ± 0.002 | 25.617 ± 0.086 | 2.277 ± 0.002 | −0.974 ± 0.086 | $-1.35^{+0.36}_{-0.16}$ ± 0.2 (syst.) | R |
| 7 | 340.432550 | −64.416440 | 20.962 ± 0.001 | 23.069 ± 0.002 | 25.215 ± 0.064 | 2.107 ± 0.002 | −1.015 ± 0.064 | $-1.44^{+0.31}_{-0.22}$ ± 0.2 (syst.) | B |
| 8 | 340.454107 | −64.428709 | 20.977 ± 0.001 | 23.111 ± 0.002 | 25.130 ± 0.061 | 2.134 ± 0.002 | −1.182 ± 0.061 | $-1.77^{+0.18}_{-0.11}$ ± 0.2 (syst.) | B |
| 9 | 340.496129 | −64.411772 | 21.011 ± 0.001 | 23.226 ± 0.003 | 25.644 ± 0.089 | 2.215 ± 0.003 | −0.904 ± 0.089 | $-1.24^{+0.32}_{-0.27}$ ± 0.2 (syst.) | R |
| 10 | 340.474934 | −64.427693 | 21.041 ± 0.001 | 23.250 ± 0.003 | 25.787 ± 0.108 | 2.209 ± 0.003 | −0.777 ± 0.108 | $-0.97^{+0.36}_{-0.29}$ ± 0.2 (syst.) | R |
| 13 | 340.466732 | −64.418331 | 21.071 ± 0.001 | 23.151 ± 0.002 | 25.193 ± 0.062 | 2.080 ± 0.002 | −1.078 ± 0.062 | $-1.64^{+0.25}_{-0.21}$ ± 0.2 (syst.) | B |

**Note.** Measurements for the full sample of stars analyzed in this work. The "Split RGB?" column indicates whether the star was used for characterizing the bifurcated RGB ("B" for the blue RGB, "R" for the red RGB) or not used at all ("0") because it did not pass the magnitude cut described in Section 4.4. Time series HST data suggests that star 22 is variable, so its metallicity measurement may not be fully reliable. A portion of the table is presented here for form and content.

(This table is available in its entirety in machine-readable form.)

**Table 5**
EMP Star ([Fe/H] < −3.0) Candidates

| Star | R.A. (deg) | Decl. (deg) | F814W (mag) | F475W (mag) | F395N (mag) | VI (mag) | CaHK (mag) | [Fe/H] (dex) | Split RGB? |
|---|---|---|---|---|---|---|---|---|---|
| 139 | 340.468741 | −64.438595 | 22.237 ± 0.002 | 23.933 ± 0.003 | 25.057 ± 0.060 | 1.696 ± 0.004 | −1.420 ± 0.060 | $-3.04^{+0.46}_{-0.43}$ ± 0.5 (syst.) | B |
| 222 | 340.487658 | −64.410618 | 22.765 ± 0.002 | 24.289 ± 0.004 | 25.172 ± 0.059 | 1.524 ± 0.004 | −1.403 ± 0.059 | $-3.35^{+0.44}_{-0.37}$ ± 0.5 (syst.) | 0 |
| 254 | 340.439639 | −64.405761 | 22.933 ± 0.003 | 24.487 ± 0.005 | 25.505 ± 0.080 | 1.554 ± 0.006 | −1.313 ± 0.080 | $-3.15^{+0.49}_{-0.51}$ ± 0.5 (syst.) | 0 |
| 297 | 340.470610 | −64.433967 | 23.129 ± 0.003 | 24.655 ± 0.005 | 25.655 ± 0.081 | 1.526 ± 0.006 | −1.289 ± 0.081 | $-3.16^{+0.43}_{-0.49}$ ± 0.5 (syst.) | 0 |
| 301 | 340.467641 | −64.404654 | 23.148 ± 0.003 | 24.641 ± 0.005 | 25.623 ± 0.084 | 1.493 ± 0.006 | −1.257 ± 0.084 | $-3.11^{+0.54}_{-0.50}$ ± 0.5 (syst.) | 0 |
| 302 | 340.504043 | −64.411705 | 23.156 ± 0.003 | 24.659 ± 0.005 | 25.627 ± 0.091 | 1.503 ± 0.006 | −1.287 ± 0.091 | $-3.15^{+0.44}_{-0.53}$ ± 0.5 (syst.) | 0 |
| 323 | 340.446389 | −64.425761 | 23.284 ± 0.003 | 24.799 ± 0.005 | 25.854 ± 0.107 | 1.515 ± 0.006 | −1.217 ± 0.107 | $-3.02^{+0.49}_{-0.56}$ ± 0.5 (syst.) | 0 |
| 329 | 340.403822 | −64.427680 | 23.310 ± 0.003 | 24.813 ± 0.005 | 25.851 ± 0.107 | 1.503 ± 0.006 | −1.217 ± 0.107 | $-3.01^{+0.50}_{-0.57}$ ± 0.5 (syst.) | 0 |

**Note.** The eight EMP ([Fe/H] < −3.0) star candidates identified in our work.

(This table is available in its entirety in machine-readable form.)

**Table 6**
MR Stars ([Fe/H] > −1.0)

| Star | R.A. (deg) | Decl. (deg) | F814W (mag) | F475W (mag) | F395N (mag) | VI (mag) | CaHK (mag) | [Fe/H] (dex) | Split RGB? |
|---|---|---|---|---|---|---|---|---|---|
| 2 | 340.446929 | −64.422353 | 20.876 ± 0.001 | 23.150 ± 0.002 | 26.046 ± 0.131 | 2.274 ± 0.002 | −0.515 ± 0.131 | >−0.83 | R |
| 10 | 340.474934 | −64.427693 | 21.041 ± 0.001 | 23.250 ± 0.003 | 25.787 ± 0.108 | 2.209 ± 0.003 | −0.777 ± 0.108 | $-0.97^{+0.36}_{-0.29}$ ± 0.2 (syst.) | R |
| 15 | 340.432114 | −64.414623 | 21.120 ± 0.001 | 23.352 ± 0.003 | 26.205 ± 0.134 | 2.232 ± 0.003 | −0.495 ± 0.134 | >−0.90 | R |
| 21 | 340.420288 | −64.416096 | 21.185 ± 0.001 | 23.394 ± 0.003 | 26.016 ± 0.123 | 2.209 ± 0.003 | −0.692 ± 0.123 | $-0.78^{+0.42}_{-0.34}$ ± 0.2 (syst.) | R |
| 27 | 340.461866 | −64.417534 | 21.289 ± 0.001 | 23.399 ± 0.003 | 25.905 ± 0.121 | 2.110 ± 0.003 | −0.659 ± 0.121 | $-0.63^{+0.38}_{-0.40}$ ± 0.2 (syst.) | R |
| 34 | 340.491319 | −64.419207 | 21.391 ± 0.001 | 23.387 ± 0.003 | 25.607 ± 0.083 | 1.996 ± 0.003 | −0.774 ± 0.083 | $-0.78^{+0.43}_{-0.36}$ ± 0.2 (syst.) | B |
| 38 | 340.444575 | −64.421917 | 21.470 ± 0.001 | 23.544 ± 0.003 | 26.256 ± 0.148 | 2.074 ± 0.003 | −0.399 ± 0.148 | >−0.60 | R |
| 39 | 340.465754 | −64.426153 | 21.427 ± 0.001 | 23.324 ± 0.004 | 25.389 ± 0.069 | 1.897 ± 0.004 | −0.781 ± 0.069 | $-0.67^{+0.40}_{-0.42}$ ± 0.2 (syst.) | B |
| 49 | 340.465457 | −64.428299 | 21.588 ± 0.001 | 23.591 ± 0.003 | 26.147 ± 0.122 | 2.003 ± 0.003 | −0.448 ± 0.122 | >−0.66 | R |
| 51 | 340.446115 | −64.413465 | 21.601 ± 0.001 | 23.617 ± 0.003 | 25.978 ± 0.115 | 2.016 ± 0.003 | −0.663 ± 0.115 | $-0.68^{+0.38}_{-0.38}$± 0.2 (syst.) | R |

**Note.** MR stars ([Fe/H] > −1.0) identified in our work.

(This table is available in its entirety in machine-readable form.)

## ORCID iDs

Sal Wanying Fu https://orcid.org/0000-0003-2990-0830
Daniel R. Weisz https://orcid.org/0000-0002-6442-6030
Nicolas Martin https://orcid.org/0000-0002-1349-202X
Alessandro Savino https://orcid.org/0000-0002-1445-4877
Michael Boylan-Kolchin https://orcid.org/0000-0002-9604-343X
Patrick Côté https://orcid.org/0000-0003-1184-8114
Andrew E. Dolphin https://orcid.org/0000-0001-8416-4093
Mario L. Mateo https://orcid.org/0000-0002-3856-232X
Jenna Samuel https://orcid.org/0000-0002-8429-4100
Nathan R. Sandford https://orcid.org/0000-0002-7393-3595